\documentclass[aps,prb,showpacs,twocolumn]{revtex4}

\usepackage{amsmath}% math extension
\usepackage{graphicx}% include figure files

\begin{document}

\title{Comparison of ion sites and diffusion paths in glasses
  obtained by\\ molecular dynamics simulations and bond valence
  analysis}

\author{Christian M\"uller,$^{1}$ Egbert Zienicke,$^{1}$ Stefan
  Adams,$^{2}$ Junko Habasaki,$^{3}$ and Philipp Maass$^{1}$}
  \email{philipp.maass@tu-ilmenau.de}
\homepage{http://www.tu-ilmenau.de/theophys2}

\affiliation{ 
   $^{1}$Institut f\"ur Physik, Technische Universit\"at
  Ilmenau, 98684 Ilmenau,Germany\\
  $^{2}$Department of Materials Science and Engineering, National
  University of Singapore, Singapore 117576\\
  $^{3}$Tokyo Institute of Technology, 4259 Nagatsuta-cho, Yokohama
  226-8502, Japan}

\date{Dated: June 21, 2006}

\begin{abstract}

  Based on molecular dynamics simulations of a lithium metasilicate
  glass we study the potential of bond valence sum calculations to
  identify sites and diffusion pathways of mobile Li ions in a glassy
  silicate network. We find that the bond valence method is not well
  suitable to locate the sites, but allows one to estimate the number
  of sites. Spatial regions of the glass determined as accessible for
  the Li ions by the bond valence method can capture up to 90\% of the
  diffusion path. These regions however entail a significant fraction
  that does not belong to the diffusion path. Because of this low
  specificity, care must be taken to determine the diffusive motion of
  particles in amorphous systems based on the bond valence method. The
  best identification of the diffusion path is achieved by using a
  modified valence mismatch in the BV analysis that takes into account
  that a Li ion favors equal partial valences to the neighboring
  oxygen ions. Using this modified valence mismatch it is possible to
  replace hard geometric constraints formerly applied in the BV
  method. Further investigations are necessary to better understand
  the relation between the complex structure of the host network and
  the ionic diffusion paths.

\end{abstract}

%66.30.Dn Theory of diffusion and ionic conduction in solids
%66.30.Hs Self-diffusion and ionic conduction in nonmetals
%82.45.Gj Electrolytes (for polyelectrolytes, see also 82.35.Rs and 82.45.Wx

\maketitle

\section{Introduction}

On a coarse grained time scale, ionic conduction in glasses is
commonly described by a thermally activated jump motion of the mobile
ions in an irregular host framework formed by a network former
(SiO$_2$, B$_2$O$_3$,...). Regions of long residence times in the
network are associated with sites, and rare events, where ions move
between neighboring sites, are associated with jumps.  The problem how
to identify and characterize sites and diffusion paths in such
non-crystalline materials poses a challenging task of current research
activities. Its solution can be expected to play a key role in
approaching a more quantitative understanding of ion transport
properties in glasses.\cite{Dieterich/Maass:2002} Methods developed in
this field should be useful also for the description of thermally
activated transport processes in other disordered systems.

In recent molecular dynamics simulations of modified network glasses
\cite{Lammert/etal:2003,Habasaki/Hiwatari:2004,Vogel:2004,Lammert/Heuer:2004,Heuer/etal:2004,Lammert/Heuer:2005,
Habasaki/Ngai:2005} sites were identified based on the number density
$\rho({\bf x})$ of the mobile ions. In some of these works, the sites
constitute those spatial regions, where $\rho({\bf x})$ exceeds a
certain threshold $\rho_\star$. The threshold value was specified by
requiring the number of these regions to be maximal (for details of
the procedure and the question which of these regions are identified
with sites, see ref.~\onlinecite{Lammert/etal:2003} and the analysis
in Sec.~\ref{sec:mdresults}). The sites form during the cooling
process in the highly viscous melt \cite{Horbach:2004} and are stable
on the time scale of the main structural rearrangements of the
network, which, below the glass transition temperature, is much larger
than the time scale needed for the mobile ions to reach the long-time
diffusion regime. For the systems
investigated,\cite{Lammert/etal:2003,Habasaki/Hiwatari:2004,Vogel:2004,Lammert/Heuer:2004,Heuer/etal:2004,Lammert/Heuer:2005,
Habasaki/Ngai:2005} the number of sites exceeds the number of ions by
6-12\% only, suggesting that a vacancy mediated hopping dynamics
\cite{vacancy-comm} should be considered in a coarse grained
description.\cite{Peibst/etal:2005,Dyre:2003} Due to the interaction
effects the ionic motion can be strongly cooperative in such
situations, as reported by one of the authors,
\cite{Habasaki/Hiwatari:2004,Habasaki/Ngai:2005} and also in
ref.~\onlinecite{Kunow/Heuer:2005}.

Although the method of determining the local number density in MD
simulations was by following the trajectories of the mobile ions and
by registering their occupation times in small cells (over
sufficiently long time periods), it should be noted that $\rho({\bf
x})$ can be viewed as an equilibrium quantity with respect to
configurations belonging to the metastable attraction basin of the
free energy reached after the cooling process. This means, that in the
absence of aging effects, which involve transitions between different
basins, $\rho({\bf x})$ can in principle be determined independently
of the ion dynamics.

Experimentally, a measurement of the local number density of mobile
ions with respect to the host structure of the network forming ions is
currently not possible. Unlike in crystals, it is already difficult to
obtain a reliable representation of the framework structure. The most
successful procedure to date is to use x-ray and neutron diffraction
data as input for a reverse Monte Carlo modeling.
\cite{Roling/Swenson:2005} Structures obtained in this way are in
agreement with the static structure factors and they fulfill certain
constraints imposed by chemical requirements. Recently this structural
modeling has been combined with bond valence (BV) sum calculations in
order to identify diffusion paths for mobile ions in network glasses.
\cite{Adams/Swenson:2000,Swenson/Adams:2003,Hall/etal:2004,Adams/Swenson:2005}
However, while for various crystalline systems it could be
demonstrated that both sites and pathways predicted by the BV method
match the results from detailed anharmonic crystal structure
analyses,\cite{Garrett/etal:1982,Brown:2002,Adams:2000a,Adams:2000b}
the potential of the BV method for predicting diffusion pathways or
sites in amorphous systems needs to be clarified.

Since ionic transport pathways in glasses cannot be directly inferred
from experiment, we test the BV method in this work by MD simulations.
The interaction potentials used in our MD simulations were derived
from ab initio Hartree-Fock self-consistent calculations and were
checked to keep the crystal structure stable under constant pressure
conditions. \cite{Habasaki:1992} This is a severe test of the quality
of the potential parameters. Many properties of ion conducting
glasses, as e.g.\ structural information obtained from scattering
experiments, vibrational spectra, values for ionic diffusion and
various thermodynamic properties are successfully reproduced by the MD
model. Using the MD simulations as a reference to test the BV method,
we nevertheless have to bear in mind that the underlying Hamiltonian
for the interactions between the ions and the way of vitrifying in the
MD simulations (e.g.\ the high cooling rates) may not give a fully
accurate representation of all properties of the real experimental
system. In any case, even if the representation with respect to the
sites and conduction pathways is not perfect, one should expect that
the BV method is still applicable, since the MD model itself presents
a valid physical system. For the model system, however, the optimal
parameters used in the BV analysis could possibly be different from
those applied to the corresponding real glass (cf.\
Sec.~\ref{sec:bv-analysis}).

Our results show that the BV paths entail the sites of the mobile ions
in the MD simulations. The BV method yields a reasonable estimate of
the number of sites for the mobile ions, but is not well suitable for
locating the sites within the pathways. In order to evaluate the
potential of the BV method with respect to diffusion pathways, the
latter have to be specified from MD simulations. Such specification
can be done based on $\rho({\bf x})$, by using the percolation
threshold $\rho_{\rm perc}$ for attributing the regions with
$\rho({\bf x})>\rho_{\rm perc}$ to the diffusion path. We will show
that the BV path reaches a sensitivity up to 56\%, i.e.\ it covers
up to 56\% of this diffusion path, and, conversely, the specificity
lie in the range 50-60\%, i.e. 50-60\% of the BV path belong to the
diffusion path. A variant of the BV method recently developed by one
of the authors,\cite{Adams:2006} allows us, by introducing a penalty
function for unfavorable asymmetric bonding situations, to achieve a
sensitivity up to 90\% and a specificity between 30-60\%.

The paper is organized as follows. After a description of the MD
simulations in Sec.~\ref{sec:mdsimulation}, we perform an
identification of the ion sites in Sec.~\ref{sec:mdresults} following
the procedure suggested in ref.~\onlinecite{Lammert/etal:2003}. We
include a discussion of the role of the grid spacing used in this
procedure and suggest an alternative criterion to finally assign the
regions with $\rho({\bf x})>\rho_\star$ to sites. The diffusion path
is determined from a percolation analysis. In
Sec.~\ref{sec:bv-analysis} we perform a BV analysis based on
time-averaged network configurations obtained from the MD simulations
and evaluate in Sec.~\ref{sec:comparison} the potential of the BV
method for identifying ion sites and the diffusion path by comparing
the results with those obtained in Sec.~\ref{sec:mdresults}. Finally,
we give in Sec.~\ref{sec:conclusion} a summary of the results and an
outlook to further research.

\section{Molecular Dynamics Simulations}\label{sec:mdsimulation}

We perform MD simulations of a Li$_2$SiO$_3$ glass in the NVE ensemble
at a temperature of 700~K with periodic boundary conditions. Newton's
equations are solved using the Verlet algorithm with time step $\Delta
t=1$\,fs. The computational domain is a cubic box of side length
$L=16.68$\,\AA\ filled with 144 lithium, 72 silicon and 216 oxygen
ions. The box size was determined by performing simulations in the NPT
ensemble at atmospheric pressure. It corresponds to material densities
that match the experimental ones within 5\% of error.

Pair potentials of Gilbert-Ida type \cite{Gilbert:1968,Ida:1976}
describe the interactions between the species:
\begin{eqnarray}
U_{ij}(r)&=& \frac{e^2}{4\pi\varepsilon_0} \, \frac{z_i z_j}{r}+ f_0
(b_i +b_j) \exp \left( \frac{a_i+a_j-r}{b_i+b_j} \right) \nonumber\\
&&{}-\frac{c_i c_j}{r^6},
\label{eq:uij}
\end{eqnarray}
where the parameters listed in table~\ref{params} have been optimized
\cite{Habasaki:1992} and shown to give good agreement with
experimental
data.\cite{Habasaki:1992,Habasaki:1995,Banhatti/Heuer:2001,Heuer/etal:2002}
The first term in eq.~(\ref{eq:uij}) is the Coulomb interaction with
effective charge numbers $z_i$ for Li, Si and O. The long-range
Coulomb interaction with the image charges in the periodically
continued copies of the simulation box is taken into account by
standard Ewald summation. A Born-Meyer type potential
$A_{ij}\exp(-r/\lambda_{ij})$ in the second term of eq.~(\ref{eq:uij})
takes into account the repulsive short-range interactions, where the
parameters $a_i$ and $b_j$ appearing in (\ref{eq:uij}) decompose
$A_{ij}$ and $\lambda_{ij}$into values assigned to the interacting
species. The last term in eq.~(\ref{eq:uij}) is a dispersive
interaction and present only for interactions between oxygen ions with
distance larger than $r_{\rm c}=1.3$\,\AA.\cite{dispersive-comm}

\begin{table}[tb]
\caption{\label{params} Potential parameters for the MD simulations
(cf.\ eq.~(\ref{eq:uij})).}
\begin{center}
\begin{tabular}
{|c||@{\hspace{1em}}c@{\hspace{1em}}|@{\hspace{1em}}c@{\hspace{1em}}|@{\hspace{1em}}c@{\hspace{1em}}|@{\hspace{1em}}c@{\hspace{1em}}|}\hline
 Ion & $z$ & $a$ [\AA] & $b$ [\AA] & $c$ [\AA$^3\sqrt{\rm kJ/mol}$] \\
 \hline \hline Li$^+$ & \phantom{-}0.87 & 1.0155 & 0.07321 &
 \phantom{1}22.24 \\ \hline Si$^{4+}$ & \phantom{-}2.40 & 0.8688 &
 0.03285 & \phantom{1}47.43 \\ \hline O$^{2-}$ & -1.38 & 2.0474 &
 0.17566 & 143.98 \\ \hline\hline
 \multicolumn{5}{|c|}{$f_0=4.186$\,kJ\AA$^{-1}$mol$^{-1}$
\hspace{1em} $r_{\rm c}=1.3$\,\AA}\\ \hline
\end{tabular}
\end{center}
\end{table}

\begin{figure}
\hspace*{0.02\textwidth}
\includegraphics[width=0.46\textwidth,clip=,]{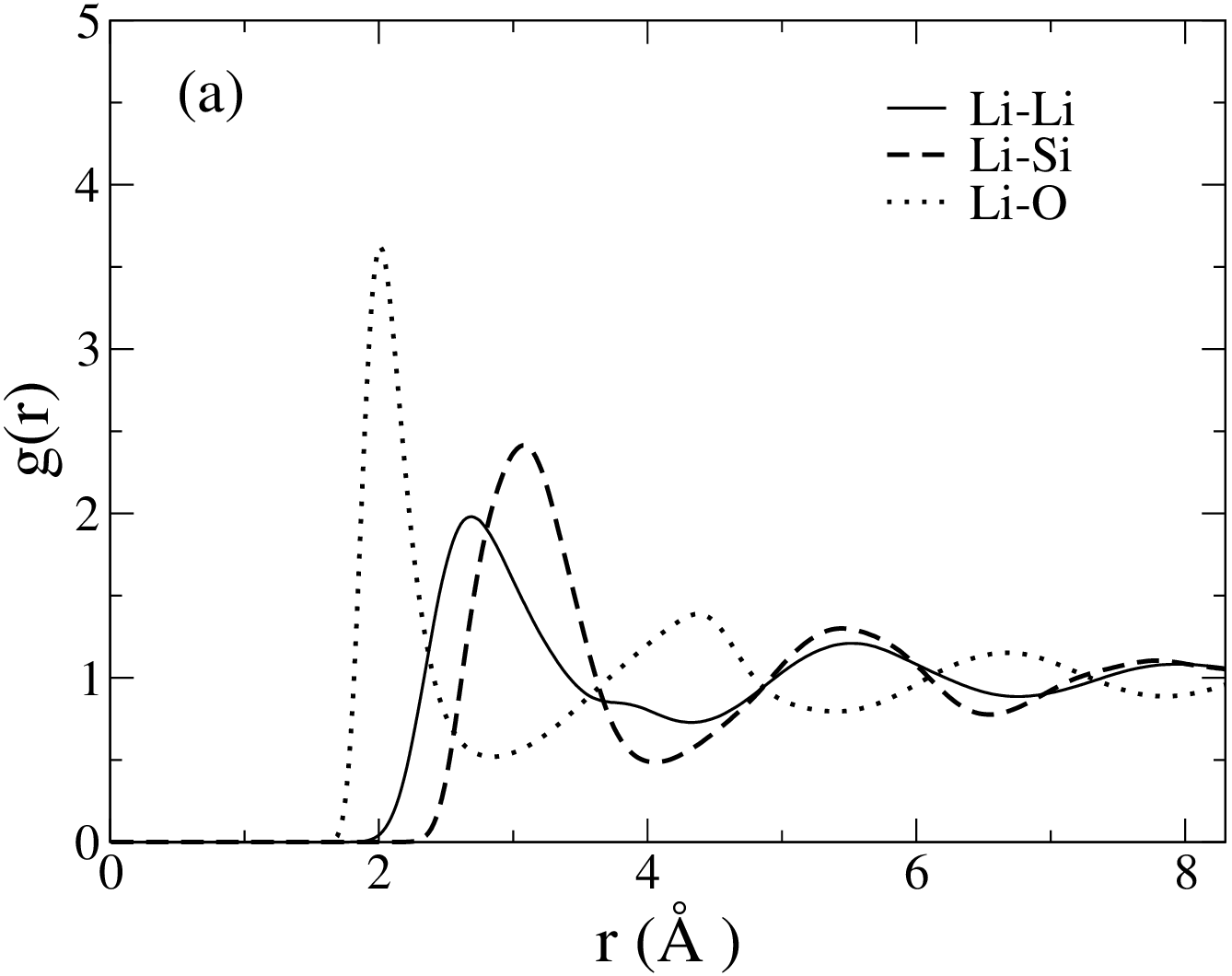}\\[3ex]
\includegraphics[width=0.48\textwidth,clip=,]{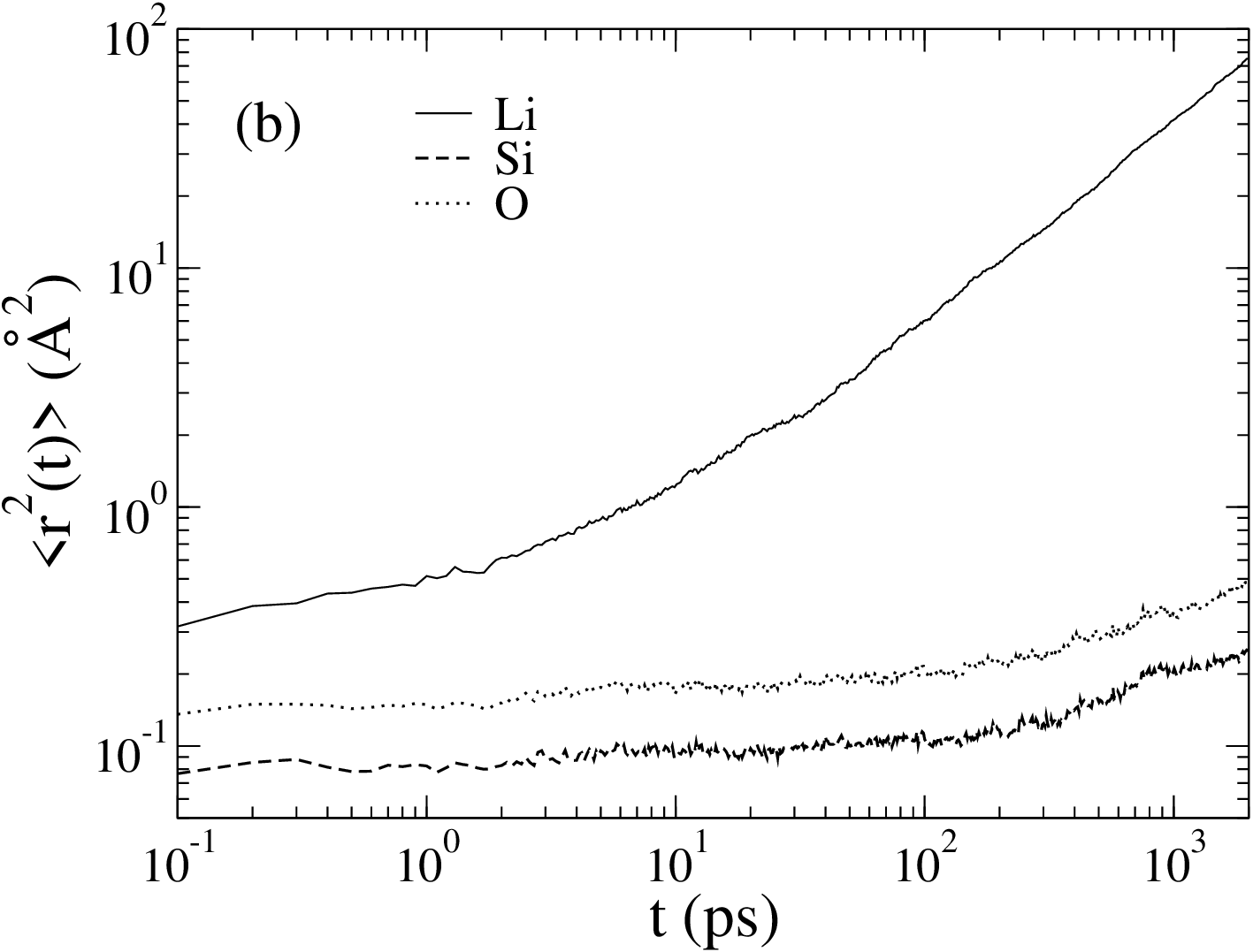}
\caption{\label{fig:gr-msd} {\it (a)} Pair correlation function of Li-Li,
Li-Si and Li-O and {\it (b)} time-dependent mean square displacements
of Li, O and Si ions at $T=700$\,K.}
\end{figure}

We cooled down the system in the NVT ensemble using a velocity scaling
to reach the final temperature of $T=700$~K and subsequently
equilibrated the system in the NVE ensemble. For the analysis of the
results we performed a simulation run over 40~ns.
Figure~\ref{fig:gr-msd}a shows the pair correlation functions for
Li-Li, Li-Si and Li-O that allow us to check if the minimal distances
between pairs of ions used in the BV analysis (2.48~\AA\ for Li-Si and
1.7~\AA\ for Li-O, see below) fit to the potential model
(\ref{eq:uij}). The time-dependent mean square displacement of Li, O
and Si ions is displayed in Figure~\ref{fig:gr-msd}b. The mean square
displacement of the Li ions becomes normal diffusive at time scales
larger than several hundred picoseconds, while that of the Si and O
ions practically stays constant over the whole time interval. On the
time scale of 2~ns, for which most of the calculations for the
determination of sites and diffusion paths are carried out, aging
effects of the network structure can be neglected.

\section{Identification of ion sites and the diffusion path}
\label{sec:mdresults}

During their motion the Li ions explore only parts of the host
network. To identify the regions encountered by the Li ions and their
favorable sites, the local number density $\rho(\mathbf{x})$ of Li
ions is calculated. Dependent on a threshold value $\rho_{\rm th}$, we
define a path $\mathcal{P}(\rho_{\rm
th})=\{\mathbf{x}|\rho(\mathbf{x})>\rho_{\rm th}\}$ as the region with
$\rho(\mathbf{x})>\rho_{\rm th}$, and perform a subsequent cluster
analysis of the paths in dependence of $\rho_{\rm th}$. To determine
$\rho(\mathbf{x})$ the simulation box is subdivided into a grid of
cubic cells with spacing $\Delta$ and the time $t_i$ is registered,
where it is occupied by a Li ion. The average density
$\rho_{i}\simeq\rho({\bf x}_i)$ in cell $i$ is then calculated as
$\rho_i=(t_i/t_{\rm sim})\Delta^{-3}$, where $t_{\rm sim}$ is the
total simulation time.

\begin{figure}
\includegraphics[width=0.48\textwidth,clip=,]{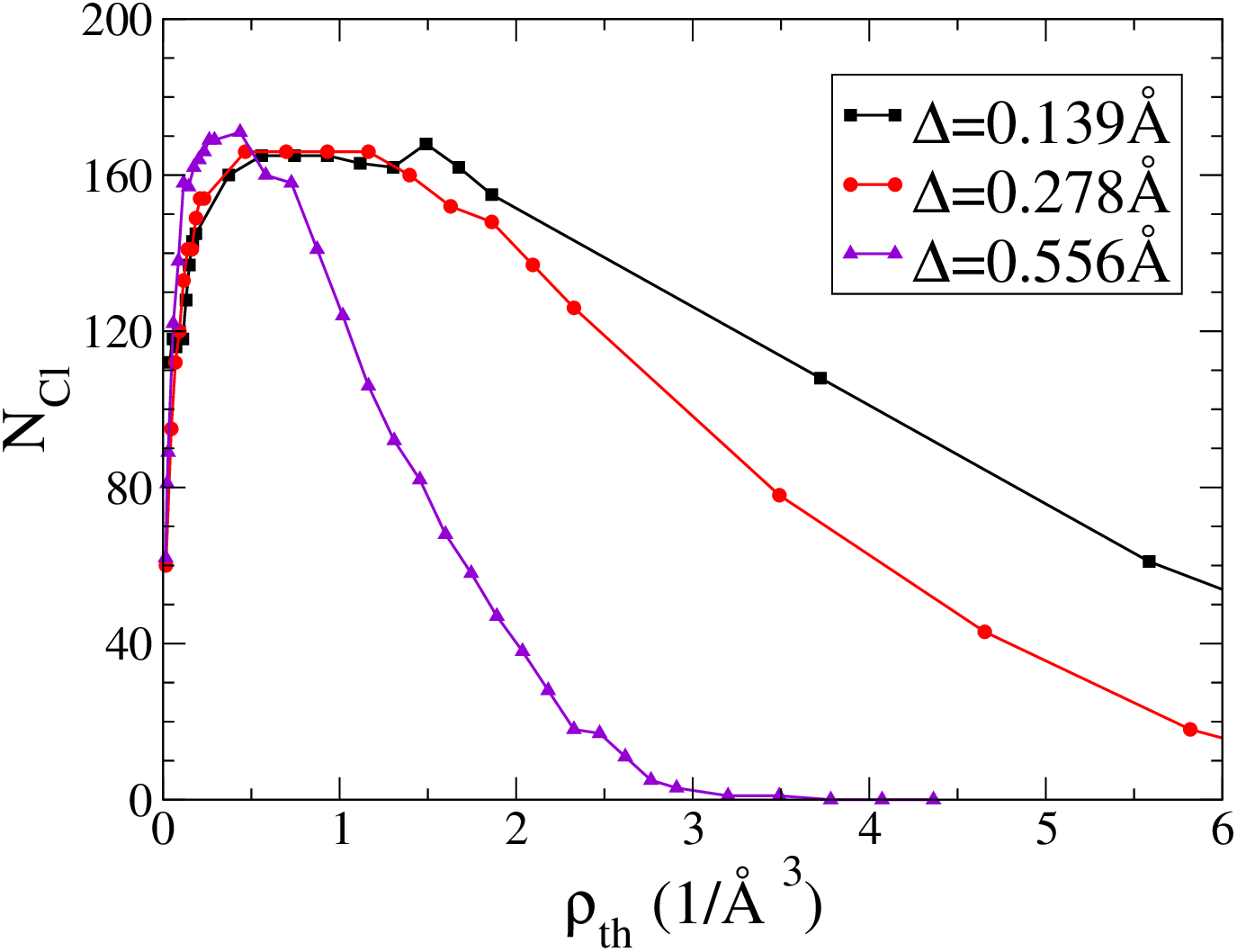}
\caption{\label{fig:nr_of_clusters} Number of clusters $N_{\rm cl}$ in
  dependence of $\rho_{\rm th}$ for three different grid spacings
  $\Delta$. Note that the mean number density of Li ions is
  $144/(16.68\mbox{\AA})^{3}\simeq0.031\mbox{\AA}^{-3}$ only, which in
  comparison with the relevant scale of $\rho_{\rm th}$ implies that
  the Li ions are strongly localized on the clusters.}
\end{figure}

In order to identify regions of high probability of occupation we
determine clusters of connected (face sharing) cells $i$ with
$\rho_i>\rho_{\rm th}$ by the Hoshen-Kopelman
algorithm.\cite{Hoshen/Kopelman:1976} Figure~\ref{fig:nr_of_clusters}
shows the number of clusters $N_{\rm cl}(\rho_{\rm th})$ in dependence
of $\rho_{\rm th}$ for three grid spacings $\Delta$. All three curves
have the same shape: Starting from large $\rho_{\rm th}$, $N_{\rm
cl}(\rho_{\rm th})$ increases with decreasing $\rho_{\rm th}$, since
an increasing number of local maxima in $\rho(\mathbf{x})$ is
identified. The strong dependence on the grid spacing at large
$\rho_{\rm th}$ shows that the local maxima in $\rho(\mathbf{x})$ are
sharp. For larger $\Delta$, less local maxima and consequently less
clusters are found, since the average density $\rho_i$ in a cell
containing a sharp local maximum of $\rho(\mathbf{x})$ becomes smaller
and can fall below $\rho_{\rm th}$. As long as $\Delta$ is much
smaller than the typical distance between the local maxima one expects
that with decreasing $\rho_{\rm th}$ eventually all local maxima are
resolved. This is indeed the case, as can be seen from the fact that
the curves for different $\Delta$ pass the same maximum at $N_{\rm
cl}^{\rm max}\simeq 165$. By further decreasing $\rho_{\rm th}$
different clusters merge and $N_{\rm cl}(\rho_{\rm th})$ decreases.

The discussion makes clear that there exists a plateau $N_{\rm
cl}(\rho_{\rm th})=N_{\rm cl}^{\rm max}$ in the range $\rho_{\rm
min}<\rho_{\rm th}<\rho_{\rm max}$, which depends on the grid spacing
(while the value $N_{\rm cl}^{\rm max}$ is not affected for
sufficiently small $\Delta$). Above $\rho_{\rm max}$ some of the local
maxima in $\rho({\bf x})$ are not resolved and below $\rho_{\rm min}$
clusters coalesce (which does not exclude that some of the $N_{\rm
cl}^{\rm max}$ clusters contain more than one local maximum of
$\rho({\bf x})$). To identify clusters with possible Li sites one
could choose any value of $\rho_{\rm th}$ in the threshold range
$\rho_{\rm min}<\rho_{\rm th}<\rho_{\rm max}$. This would not change
the identity of the possible sites but their size. In order to cover
as much volume as possible we use $\rho_\star\equiv\rho_{\rm
min}\simeq0.56$\,\AA$^{-3}$ for a unique definition of the clusters,
which are candidates for sites (strictly speaking, one should use the
value $\rho_{\rm min}$ in the limit $\Delta\to0$, since $\rho_{\rm
min}$ depends weakly on $\Delta$). The subsequent analysis is carried
out for the smallest spacing $\Delta=0.139$.

\begin{figure}
\includegraphics[width=0.48\textwidth,clip=,]{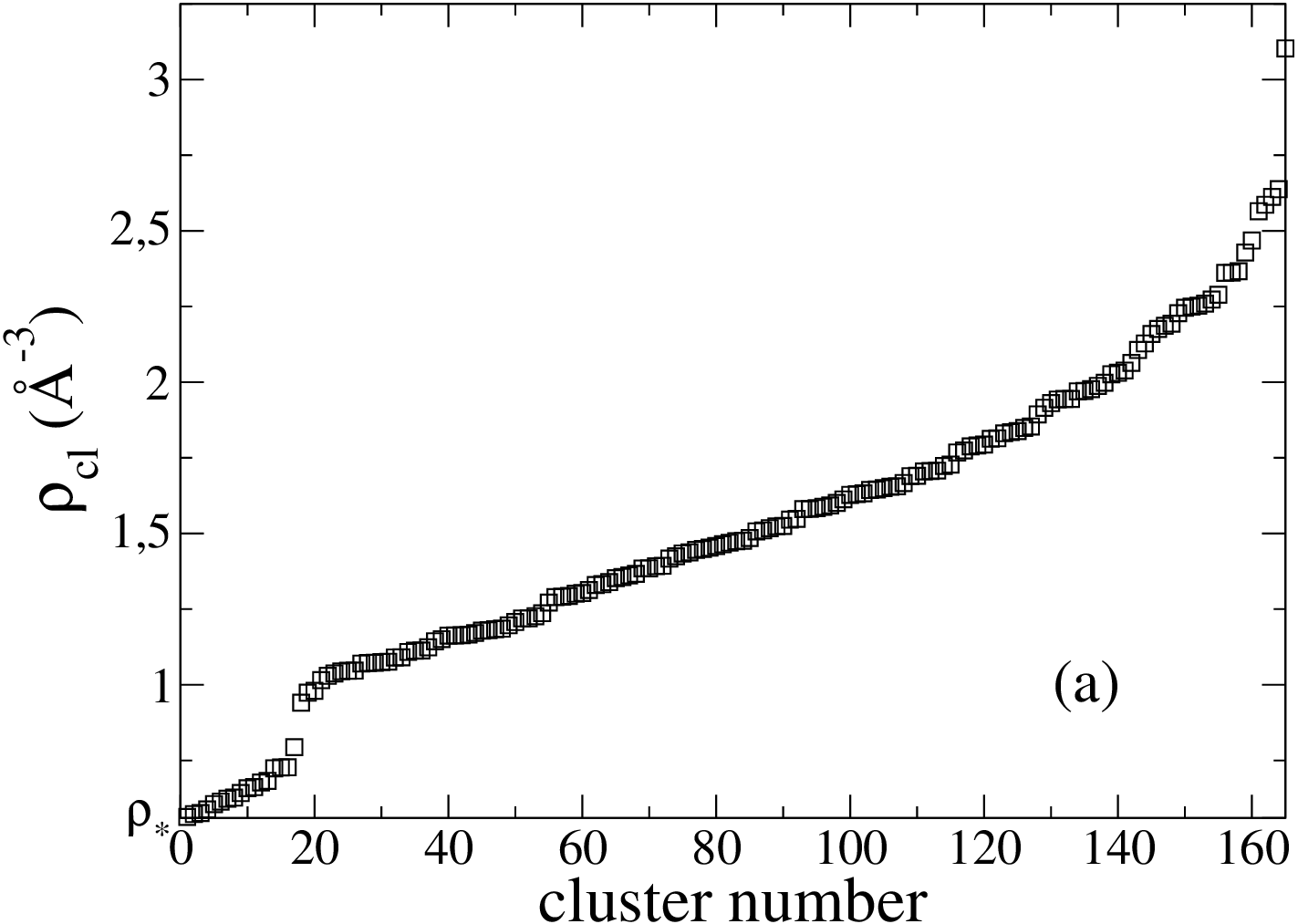}\\
\vspace*{3ex}
\includegraphics[width=0.48\textwidth,clip=,]{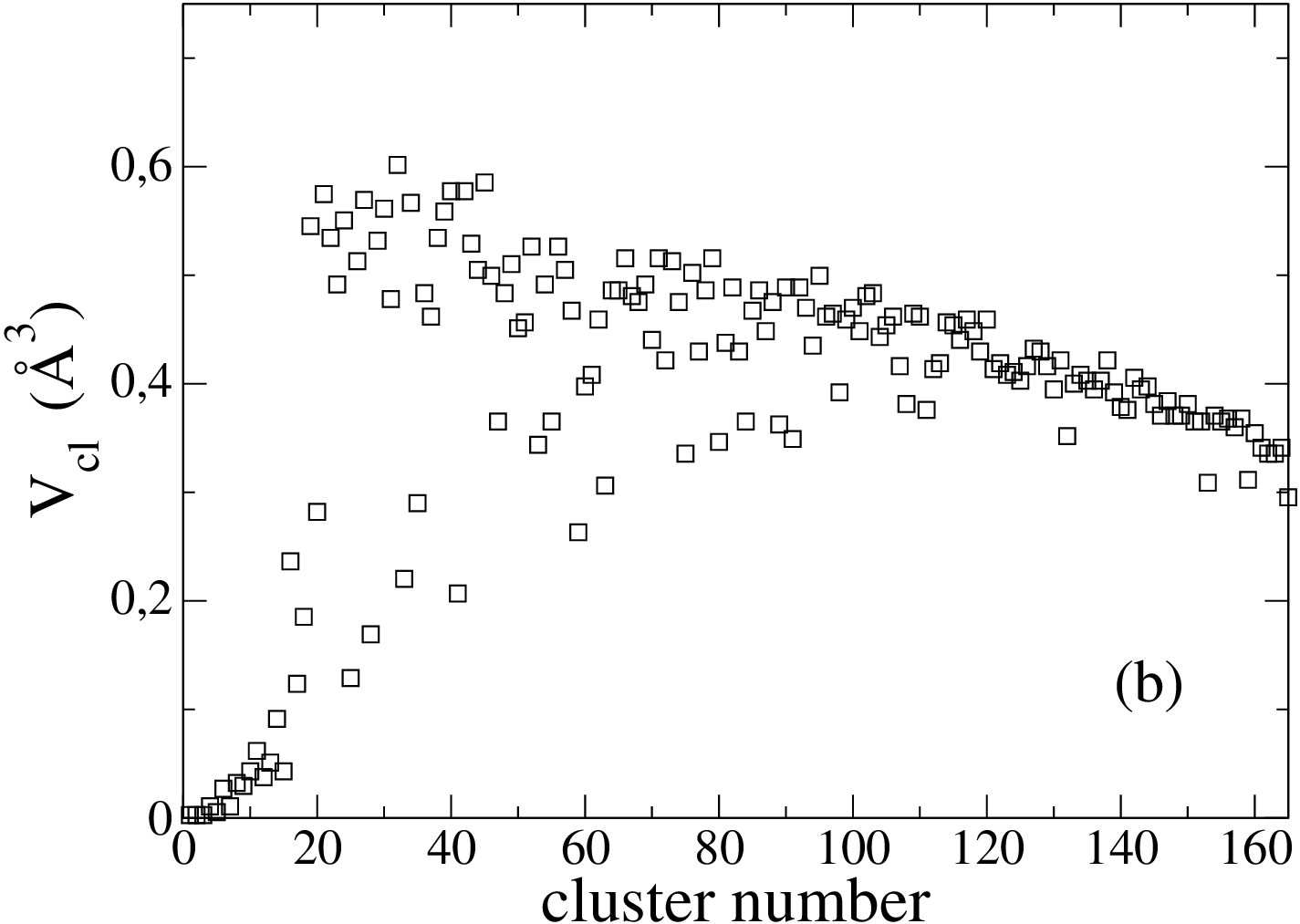}\\
\vspace*{3ex}
\includegraphics[width=0.48\textwidth,clip=,]{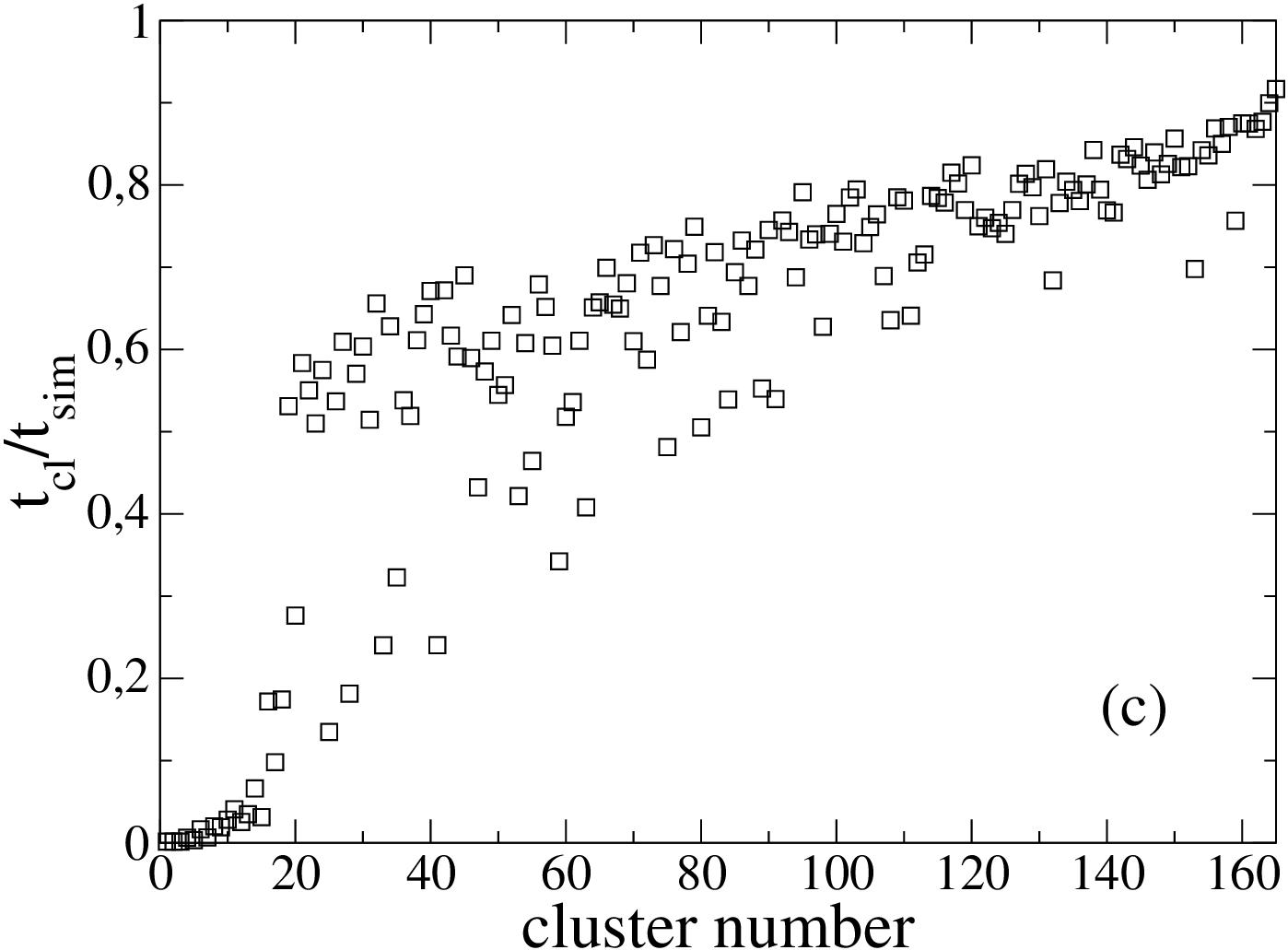}
\caption{\label{fig:cluster-properties} {\it (a)} Mean number density 
  $\rho_{\rm cl}$ of Li ions on the 165 clusters identified by the
  Hoshen-Kopelman algorithm; {\it (b)} Volume $V_{\rm cl}$ and {\it
    (c)} Li occupation probability $t_{\rm cl}/t_{sim}$ of the
  clusters.  The clusters are sorted according to increasing
  $\rho_{\rm cl}$.}
\end{figure}

Next we study properties of the clusters with respect to their
assignment to sites. To this end we determine the volume $V_{\rm
cl}=n_{\rm cells}\Delta^3$ of the clusters and the mean number density
$\rho_{\rm cl}=\sum_{i\in{\rm cl}}\rho_i/n_{\rm cells}$ of Li ions on
them, and number the clusters in order of increasing $\rho_{\rm cl}$.
$n_{\rm cells}$ denotes the number of cells in the corresponding
cluster. Figures~\ref{fig:cluster-properties}a,b,c show $\rho_{\rm
cl}$, $V_{\rm cl}$, and the occupation probability $t_{\rm cl}/t_{\rm
sim}= \sum_{i\in{\rm cl}}t_i/t_{\rm sim}=\rho_{\rm cl}n_{\rm
cells}\Delta^3$ for the numbered clusters. The nearly constant
increase of the density is interrupted by a jump after cluster number
17. Also the other two quantities show a strong increase in this
region. The first clusters are very small and have very low occupation
times. After the jump, the volume and the occupation time lie on a
main branch between 0.6 and 0.35 \AA$^3$ for the volume and between 50
and 90\% of the ratio of occupation time to simulation time. In
between there are some outliers with less volume or less occupation
time, but above the values of the first 17 clusters. With these
results one faces two problems: {\it (i)} Summing up the occupation
probability of all clusters, one finds that the total occupation
probability $\sum_{\rm cl}t_{\rm cl}/t_{\rm sim}$ of all clusters is
70\% only. {\it (ii)} According to the behavior of all three shown
quantities --- very small occupation probability, volume and
occupation time --- the first clusters before the jump do not fit to
the physical picture which one has of a site.

Problem {\it (i)} can be solved as follows. The criterion $\rho({\bf
x})>\rho_\star$ cuts off the outer fringe of a site, which is visited
only a short time by a Li ion, but dynamically can be assigned to the
site, since a Li ion in general returns to the cluster after entering
its fringe. The clusters thus can be viewed to form the core of the
sites. Problem {\it (ii)} is more subtle. As suggested in
ref.~\onlinecite{Lammert/etal:2003}, one can require that a cluster
should only be assigned to a site if the mean occupation of Li ions on
it exceeds a certain threshold. However, there can exist sites, which
are visited only rarely but which still conform to the requirement of
a sufficiently large ratio between the residence time of a Li ion on
the site and the time for its transition to a neighboring site.

We therefore use a new criterion for the assignment of the clusters to
sites by associating two times with each cluster, the total residence
time $t_{\rm res}$ and the total hopping time $t_{\rm hop}$ to any of
its neighbors. The hopping time is calculated as follows: When a Li
ion leaves a cluster and enters a new one without returning, it
performs a transition between two clusters. The hopping time then is
the time interval between the departure from the initial cluster and
the arrival at the target cluster. Summing over all transitions
starting from a given cluster, one obtains the total hopping time from
this cluster. The time interval between the entrance of a Li ion into
a cluster and the onset of its transition to another cluster yields
the residence time, and by taking the sum over all events we obtain
$t_{\rm res}$.\cite{time-comm} Note that according to this definition
the residence time includes events, where a Li ion leaves the cluster
but returns to it before entering another cluster.\cite{rattling-comm}

\begin{figure}
\includegraphics[width=0.48\textwidth,clip=,]{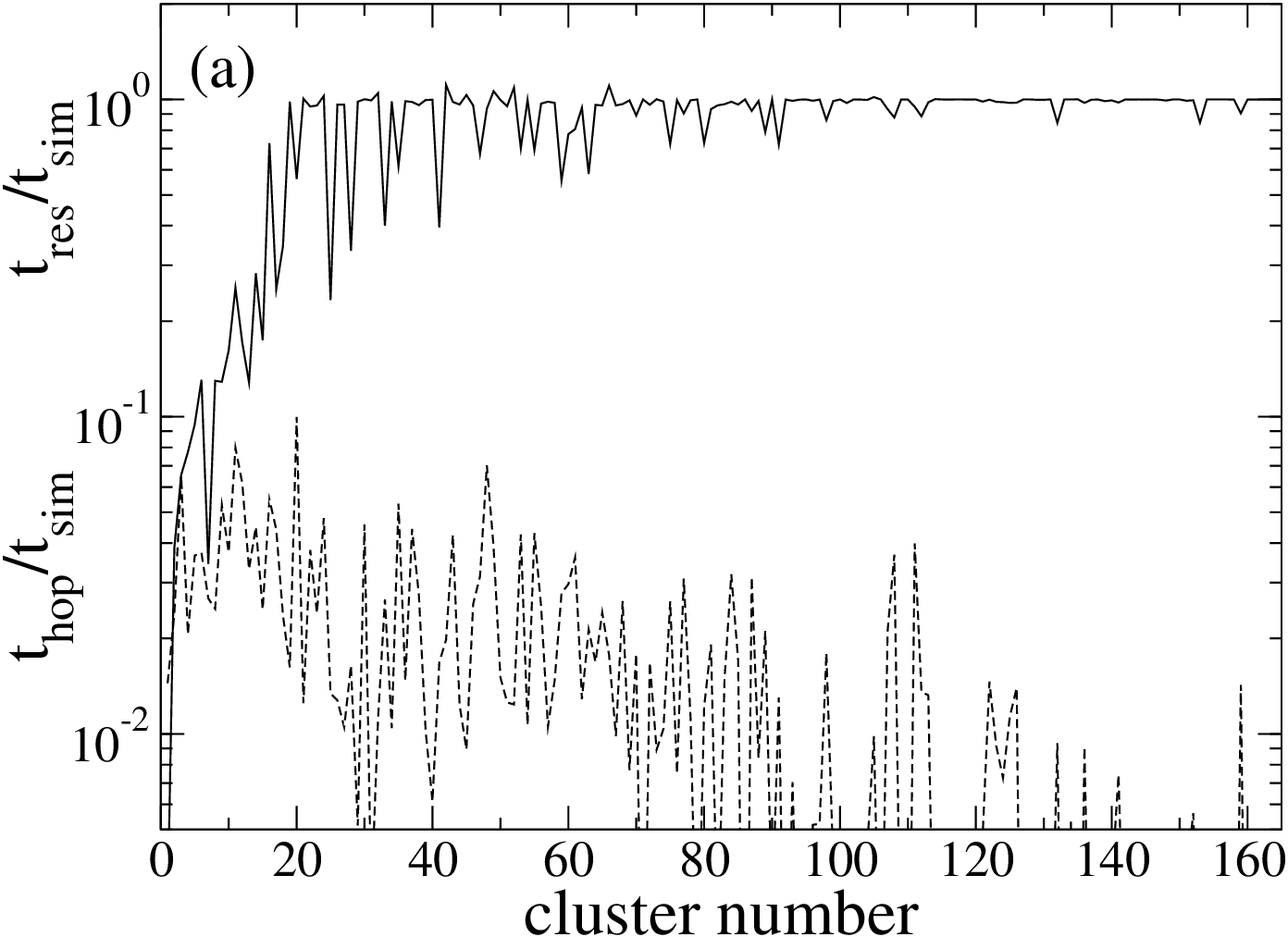}\\
\vspace*{3ex}
\includegraphics[width=0.48\textwidth,clip=,]{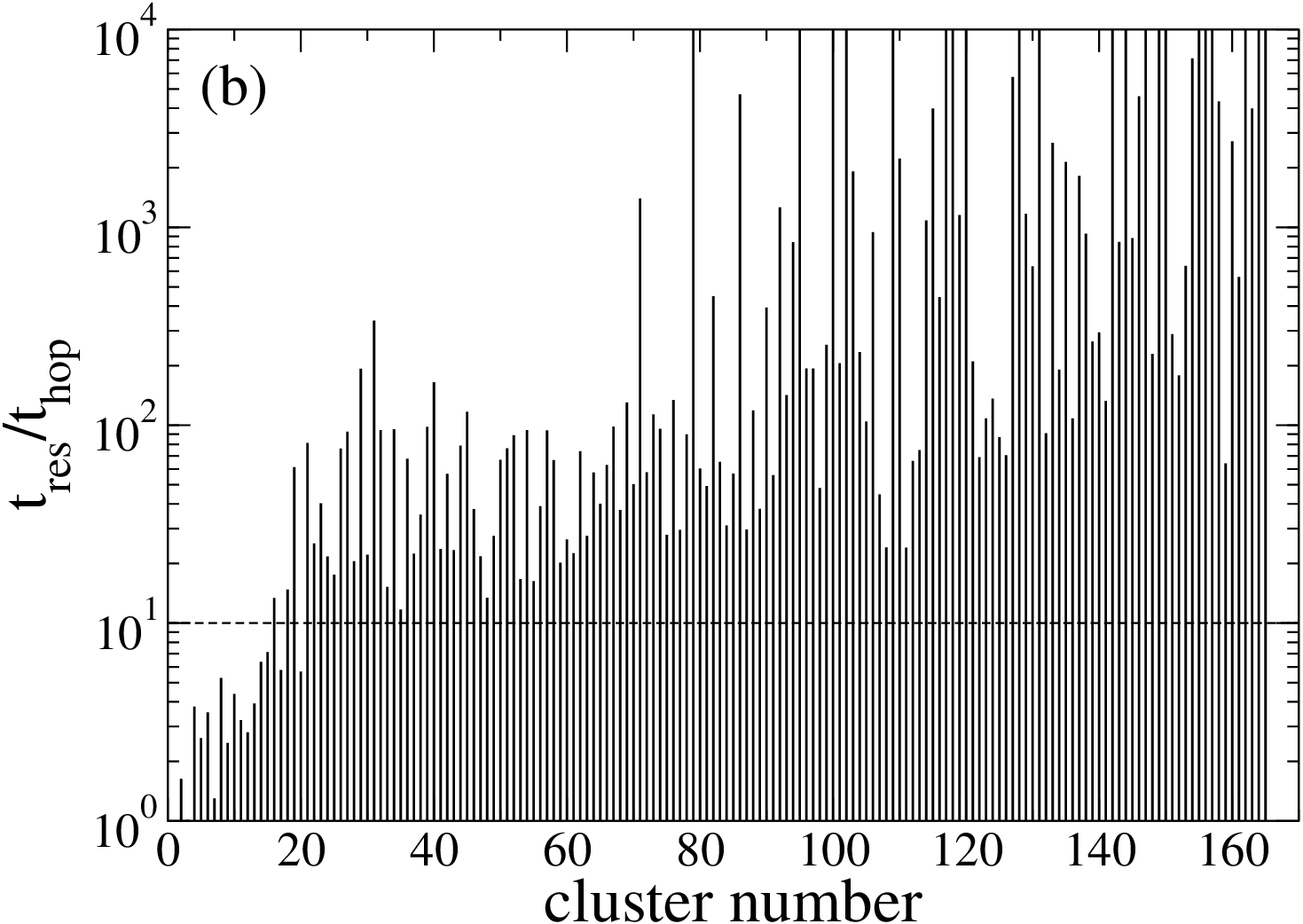}
\caption{\label{fig:res-hop-time} {\it (a)} Total residence and hopping time
  normalized with respect to the simulation time $t_{\rm sim}$, and
  {\it (b)} their ratio for the 165 clusters determined by the
  Hoshen-Kopelman algorithm.  The dashed line marks the threshold
  above which clusters are identified as sites (cf.\ 
  eq.~\ref{eq:site-cond}).}
\end{figure}

Figure~\ref{fig:res-hop-time}a shows $t_{\rm res}$ and $t_{\rm hop}$
for the numbered clusters. While $t_{\rm hop}$ fluctuates from cluster
to cluster but never exceeds one tenth (0.2~ns) of the simulation time
(2~ns), $t_{\rm res}$ shows a more smooth variation with the cluster
number as a consequence of a strong correlation to $\rho_{\rm cl}$. In
particular, clusters with the smallest $\rho_{\rm cl}$ correspond to
the one with lowest $t_{\rm res}$. The total residence probability
$\sum_{\rm cl}t_{\rm res}/t_{\rm sim}$ of the Li ions on the clusters
is now 98\%, which confirms that the problem with the low total
occupation probability $\sum_{\rm cl}t_{\rm cl}/t_{\rm sim}$ is
resolved by taking into account the fringes of the clusters.

In order for a cluster to be assigned to a site, we now use the
criterion that its residence time $t_{\rm sim}$ should be at least one
order of magnitude larger than its total hopping time $t_{\rm
hop}$,\cite{site-definition-comm}
\begin{equation}
\frac{t_{\rm res}}{t_{\rm hop}}>10\,.
\label{eq:site-cond}
\end{equation}
Figure~\ref{fig:res-hop-time}b shows $t_{\rm res}/t_{\rm hop}$ for the
numbered clusters. Most clusters fulfill condition
(\ref{eq:site-cond}), which demonstrates that the Li motion can be
represented by a hopping dynamics on a coarse grained time scale
larger than $t_{\rm hop}$. Only 17 fail to satisfy condition
(\ref{eq:site-cond}), and almost all of those have small numbers
corresponding to low $\rho_{\rm cl}$. This shows that there is hardly
any difference here between a criterion based on a threshold number
density and (\ref{eq:site-cond}). Accordingly, the sites are
determined essentially by the equilibrium quantity $\rho({\bf x})$ (in
the metastable basin after the cooling when disregarding slow
non-equilibrium aging processes of the network structure). In total,
we find 148 sites, which are only 2.8\% more than the number of Li
ions. This fraction of empty sites is slightly lower than that found
in earlier studies
\cite{Lammert/etal:2003,Habasaki/Hiwatari:2004,Vogel:2004} and
supports the picture of an ion hopping with a small number of
vacancies. Different from the usual situation in crystalline systems
in thermal equilibrium, the concentration of vacancies here should be
considered as a result of the freezing process and will not change
significantly with temperature below the glass transition.

After having identified the sites, we determine the diffusion path. To
this end we evaluate the percolation threshold\cite{Bunde/Havlin:1996}
$\rho_{\rm perc}$ and the corresponding subset $\mathcal{P}_{\rm
perc}=\mathcal{P}(\rho_{\rm perc})$ of cells, on which the Li ions can
diffuse across the system. We find $\rho_{\rm
perc}=0.030\,\mbox{\AA}^{-3}$, where $\mathcal{P}(\rho_{\rm perc})$
covers 7.0\% of the volume of the system (cf.\
table~\ref{tab:volume-fractions}). Confining the Li motion to exactly
the critical percolation path at $\rho=\rho_{\rm perc}$ would yield
anomalous sub-diffusion. However, in our case the percolation
threshold is not sharp due to the finite system size and the number
density of the Li ions takes into account only the mean occupation
probabilities of the ions but not their thermal fluctuations. Even
more important, these issues are not of particular relevance here,
since we are interested in the spatial overlap with the BV paths (see
Sec.~\ref{sec:bv-analysis}). With respect to this overlap the
criticality plays no decisive role (since connectivity properties are
not important for the overlap).

\section{Bond valence analysis}
\label{sec:bv-analysis}

The development of the bond valence approach has its origin in the search of
correlations between bond lengths, chemical valence and binding energies for
the chemical bond,\cite{Pauling:1967} and has been widely applied to crystals
with covalent and ionic bonds. The possibility to approximately describe both
types of bonding using the same formalism makes it particularly useful for
inorganic compounds with partially covalent bonds. A review of the BV method
is given in refs.~\onlinecite{Brown:2002,Adams:2000a}. We describe the method
here in connection with our application, which is the determination of sites
and diffusion paths for the Li ions based on the knowledge of the Si-O network
structure. For a Li ion to be accommodated in some region, its valence should
be close to its ``natural values'', and additional constraints for its
coordination number and minimal distance to neighboring ions are to be
satisfied (see below).

According to Pauling,\cite{Pauling:1967} each bond in a structure
induces, due to its polarity, bond valences of opposite sign (partial
charges in units of the elementary charge $e$) at the two atoms that
it connects. In an equilibrated structure, the bond valences of an ion
induced by neighboring counterions should add to its ideal value
(i.e.\ +1 for a Li ion). Since the effective overlap of electronic
orbitals typically decreases exponentially with distance of the atomic
nuclei, the partial valence $s_j({\bf x})$ of a Li ion at position
${\bf x}$ induced by a oxygen ion $j$ at position ${\bf x}_j$ is
determined by\cite{Donnay/Allmann:1970}
\begin{equation}
s_j({\bf x})=\exp\left[\frac{r_0-|{\bf x}-{\bf x}_j|}{\xi} \right],
\label{eq:sj}
\end{equation}
where $r_0$ is the ideal bond length and $\xi$ is the so-called softness
parameter that determines how fast the bond valence varies with
distance. The total valence $V({\bf x})$ of a Li ion is
computed as the sum of the bond valences $s_j({\bf x})$:
\begin{equation}
V({\bf x})=\sum_j s_j({\bf x})=\sum_j\exp\left[\frac{r_0-|{\bf x}-{\bf
x}_j|}{\xi} \right]\,.
\label{eq:v}
\end{equation}
Because of the exponential decay with distance the sum can be extended
over all oxygen ions in the computational domain (taking into account
the minimum image convention for periodic boundary conditions). The
decomposition (\ref{eq:v}) into bond valences allows one also to
associate a coordination number $C({\bf x})$ of a Li number at
position ${\bf x}$. This is defined by the number of oxygen ions
contributing $s_j({\bf x})$ exceeding a threshold $s_{\rm min}$,
\begin{equation}
C({\bf x})=\sum_j\theta\left(s_j({\bf x})-s_{\rm min}\right)\,,
\label{eq:c}
\end{equation}
where $\theta(.)$ is the Heaviside jump function ($\theta(x)=1$ for
$x>0$ and zero else).\cite{Adams/Swenson:2005} Here we choose
$s_{\rm min}$ as suggested by Brown.\cite{Brown:2002}
The BV parameters $r_{0}$,
and $\xi$ are derived from a variety of crystalline
phases,\cite{bv-param,Adams:2001} and are given in
table~\ref{tab:bv-parameter}.

\begin{table}[tb]
\caption{\label{tab:bv-parameter} Parameters used in the BV analysis.}
\begin{center}
\begin{tabular}
{|c|c|c|c|c|c|c|c|}\hline
$\xi$ [\AA] &
$r_0$ [\AA] & 
$s_{\rm min}$ &
$C_{\rm min}$ &
$C_{\rm max}$ &
$R_{\rm LiO}$ [\AA] &  
$R_{\rm LiSi}$ [\AA] &  
$\bar V$\\ \hline
0.516 & 1.17096 & 0.04 & 4 & 6 & 1.70 & 2.48 & 0.814\\ \hline
\end{tabular}
\end{center}
\end{table}

\begin{figure}
\includegraphics[width=0.48\textwidth,clip=,]{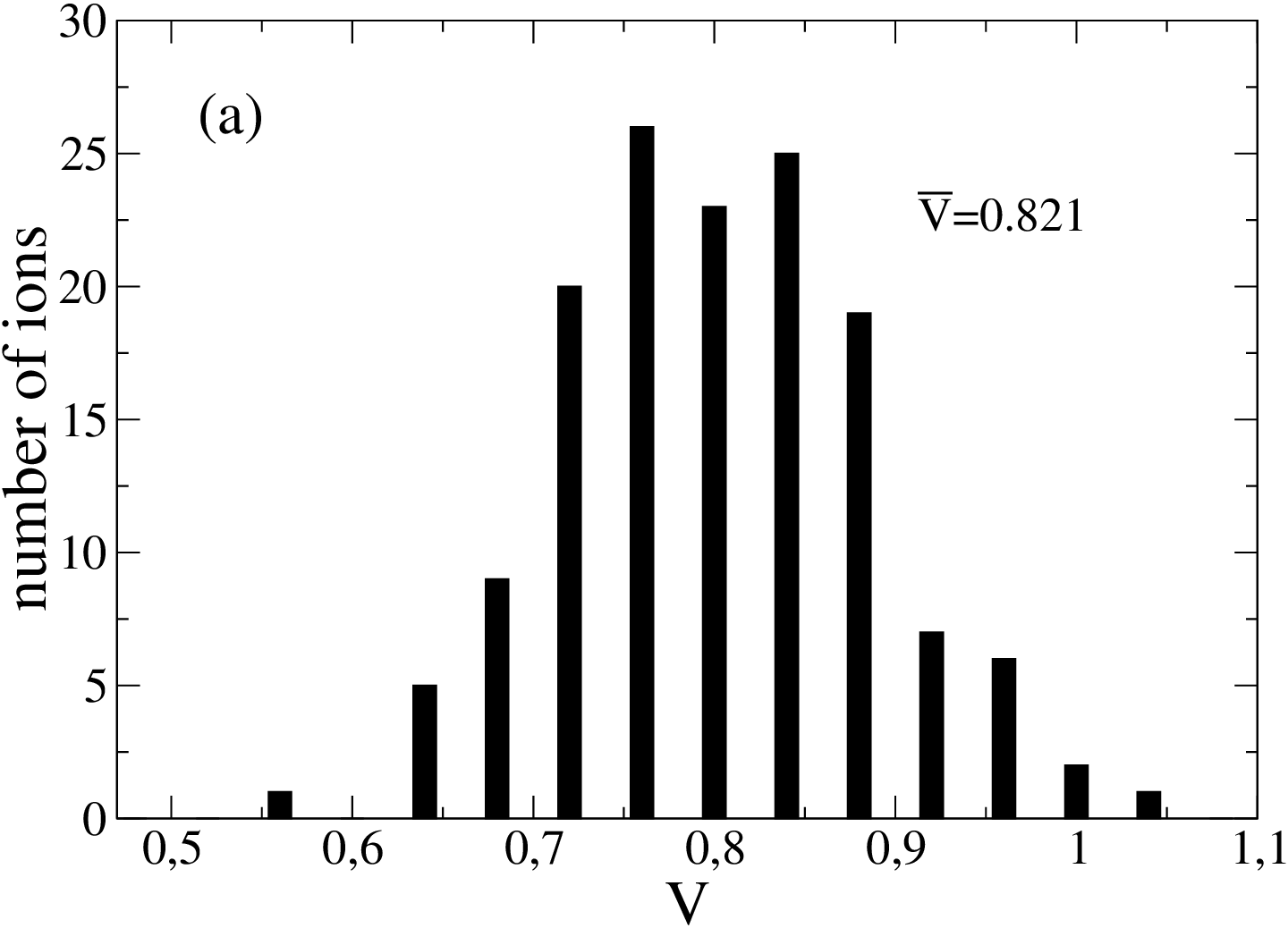}
\includegraphics[width=0.48\textwidth,clip=,]{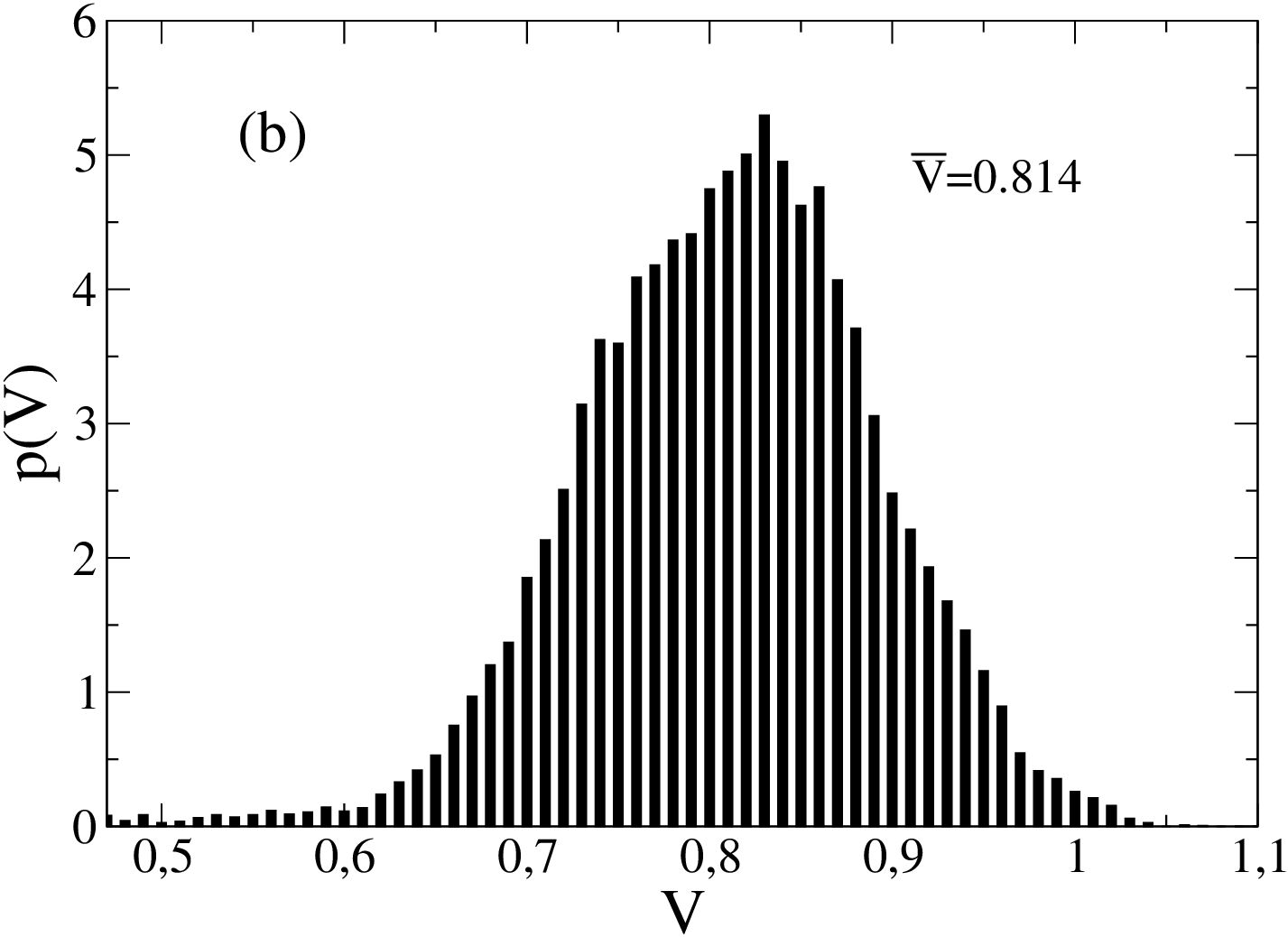}
\caption{\label{fig:valence_sum} {\it (a)} Histogram of 
  of bond valence sums for an instantaneous configuration of
  the equilibrated Li$_2$SiO$_3$ glass; {\it (b)} Probability density of bond
  valence sums with respect to the time-averaged structure of network
  forming O and Si ions.}
\end{figure}

Figure~\ref{fig:valence_sum}a shows the histogram of valences of the
144 Li ions for an instantaneous equilibrated configuration. While in
a crystalline structure the Li ions $i$ at their equilibrium positions
${\bf x}_i$ have valences $V_i=V({\bf x}_i)$ close to the ideal value
$V_{\rm id}=1$, we find a mean value $\bar V_{\rm inst.}=0.821$
significantly smaller than $V_{\rm id}$ in the (instantaneous)
amorphous glass structure. An apparent under-bonding of the Li ions
with average BV sums of ca.\ 0.9 has been found also in the BV
analysis of RMC models obtained from scattering
data\cite{Adams/Swenson:2005} and can be traced back to the
non-crystalline structure of the glass. Since in the MD simulation the
cooling rates are much larger than in experiments, the deviation from
the ideal bonding situation in a crystal at thermal equilibrium is
even larger.

With respect to the application of the BV analysis to structural
models obtained from RMC simulations, one should take into account
that the information provided by x-ray and neutron diffraction data
corresponds to time-averaged positions of the network forming ions.
This is due to the fact that the time for obtaining an evaluable
signal is orders of magnitudes larger than a typical vibrational time.
This means that a (non-unique) network configuration obtained in RMC
modeling is a representative for a time-averaged density. We thus
determined the mean positions of oxygen and silicon ions in a time
interval of 2\,ns and calculated the valences of Li ions with respect
to these mean positions. For a Li ion placed at the centers ${\bf
x}_i$ of the cells $i$ described in the previous
section~\ref{sec:mdresults} (spacing $\Delta=0.139$\,\AA), the
valences $V_i=V({\bf x}_i)$ were calculated. Taking into account the
probability of occupation $\rho({\bf x}_i)\Delta^3$ of the cells, we
then determined the probability density $p(V)$ of valences shown in
Fig.~\ref{fig:valence_sum}b. Its mean $\bar V=0.814$ is even slightly
smaller than that for the instantaneous configuration, confirming that
the ideal value $V_{\rm id}=1$ is not the preferred one in the MD
simulation of an amorphous glass structure. A small asymmetry is seen
in $p(V)$ with a steeper decrease from its maximum on the side of
large valences $V$. This reflects the asymmetry of the two-body
interaction potentials of Li with O ions, where small distances
corresponding to the repulsive part yield larger valences according to
eq.~(\ref{eq:sj}), while large distances corresponding to the
attractive part yield smaller valences. Since the asymmetry is not
pronounced, the mean value $\bar V$ is close to the value where $p(V)$
attains its maximum.

In the following we analyze the data with respect to the time-averaged
network structure and consider $\bar V$ as the ``optimal value'' of
the bond valence sum. The valence mismatch of a Li ion at position
${\bf x}$ is then given by
\begin{equation}
d({\bf x})=|V({\bf x})-\bar V|\,.
\label{eq:d}
\end{equation}
The mean valence mismatch 12\% calculated from $p(V)$ in
Fig.~\ref{fig:valence_sum}b is significantly larger than in
crystalline structures (where typical values are less than 5\%).

Following the BV method, a position ${\bf x}$ is accessible for a Li
ion if the following conditions are fulfilled:

\begin{itemize}

\item[{\it (i)}] The distances of the Li cation to silicon cations
must exceed a minimum distance
\begin{equation}
\min_j\{|{\bf x}-{\bf x}_j^{\rm Si}|\}>R_{\rm LiSi}\,,
\label{eq:cond1}
\end{equation}
$R_{\rm LiSi}=2.48$~\AA\ is chosen to be equal to (or a little less
than) the sum of the radii of the two ions in agreement with the
correlation hole of the pair correlation shown in
Fig.~\ref{fig:gr-msd}a.

\item[{\it (ii)}] According to the ``equal valence
rule''\cite{Brown:2002} among the conceivable states with matching BV
sum, the one with more symmetric bonds is energetically preferable.
The simplest way of excluding that a matching BV sum is achieved by an
unphysical strong asymmetric coordination shell is to define a minimum
acceptable bond distance,
\begin{equation}
\min_j\{|{\bf x}-{\bf x}_j^{\rm O}|\}>R_{\rm LiO}\,.
\label{eq:cond2}
\end{equation}
A value of $R_{\rm LiO}=1.7$~\AA\ is in agreement with the correlation
hole in Fig.~\ref{fig:gr-msd}a and restricts a partial Li-O bond
valence to values smaller than 0.36.

\item[{\it (iii)}] The coordination number has to lie between a
minimal and maximal value,
\begin{equation}
C_{\rm min}\le C({\bf x})\le C_{\rm max}\,,
\label{eq:cond3}
\end{equation}
where $C_{\rm min}=4$ and $C_{\rm max}=6$. This condition has
previously been assumed for sites only.\cite{Adams/Swenson:2005}

\item[{\it (iv)}] The valence mismatch must be smaller than a
  threshold value $d_{\rm th}$,
\begin{equation}
d({\bf x})=|V({\bf x})-\bar V|<d_{\rm th}\,.
\label{eq:cond4a}
\end{equation}

\end{itemize}

As an alternative to the three conditions {\it (ii-iv)} one can use
just one condition {\it (v)}, which is a modification of {\it (iv)}.
It takes into account that a Li ion is better accommodated to the local
network environment at position ${\bf x}$ if the valences $s_j({\bf
x})$ are more symmetrically distributed among the neighboring oxygen
ions. The effect can be described by defining an ``ideal partial
valence'' $s_{\rm id}$ by $C_{\rm min}s_{\rm id}=V_{\rm
id}$,\cite{Adams/Swenson:2005,Adams:2006} corresponding to a Li ion
that, when symmetrically connected to $C_{\rm min}$ oxygen ions, has
the ideal total valence $V_{\rm id}$. The deviation from the symmetric
situation is quantified by the penalty function
\begin{equation}
      p({\bf x})=\sqrt{\sum_j \left(\frac{s_j({\bf x})}{s_{\rm
      id}}-1\right)^{2\mu}}\,,
\label{eq:penalty-function}
\end{equation}
where the sum is taken over all oxygen ions and we choose $\mu=3$. A
modified valence then is defined by $d'({\bf x})=d({\bf x})+p({\bf
x})$ and we require:
\begin{itemize}
\item[{\it (v)}] The modified valence mismatch must be smaller than a
  threshold value $d_{\rm th}$,
\begin{equation}
d'({\bf x})=|V({\bf x})-\bar V|+p({\bf x})<d_{\rm th}.
\label{eq:cond4b}
\end{equation}

\end{itemize}

When applying eq.~(\ref{eq:cond4b}) instead of eq.~(\ref{eq:cond4a})
conditions {\it (ii)} and {\it (iii)} are no longer needed. As a
consequence, instead of the four parameters $R_{\rm LiO}$, $C_{\rm
min}$, $C_{\rm max}$, and $s_{\rm min}$ the parameter $\mu$ enters the
calculation.

To perform the BV analysis, the computational domain is divided into a
grid of cells with spacing $\Delta$ analogous to the procedure used in
sec.~\ref{sec:mdresults} ($\Delta=0.139$~\AA). A cell is accessible if
a Li ion placed at its center fulfills the requirements {\it (i-iv)},
or {\it (i,v)} in the modified BV method.\cite{sign-comm} The union of
such cells is the BV path $\mathcal{P}_{\rm BV}(d_{\rm th})$ (for
conditions {\it (i-iv)}) or the BV path $\mathcal{P}_{\rm BV}'(d_{\rm
th})$

\section{Comparison of ion sites and diffusion paths}
\label{sec:comparison}

We first clarify the relevance of the purely geometric constraints
{\it (i,ii)} in combination with the coordination number condition
{\it (iii)}, without taking into account the bond valence condition
{\it (iv)}. Accordingly we distinguish between the subsets defined by
applying the constraints separately: $\mathcal{P}^{\rm BV}_{\rm ex}$
is the subset given by the geometric exclusions {\it (i,ii)}, while
$\mathcal{P}^{\rm BV}_{\rm cn}$ refers to the condition imposed solely
on the coordination number {\it (iii)}. The subset defined by the
combined conditions {\it (i-iii)} is denoted as $\mathcal{P}^{\rm
BV}_{\rm ex+cn}$.
 
\begin{table}[tb]
\caption{\label{tab:volume-fractions} Volume fractions (\%) of the
  various subsets specifying sites and diffusions paths.}
\begin{center}
\begin{tabular}
{|c|c||c|c|c|}\hline
$\mathcal{P}_{\rm sites}$ &  
$\mathcal{P}_{\rm perc}$ &
$\mathcal{P}^{\rm BV}_{\rm ex}$ &
$\mathcal{P}^{\rm BV}_{\rm cn}$ &
$\mathcal{P}^{\rm BV}_{\rm ex+cn}$ \\ \hline 
1.4 & 7.0 & 10.0 & 83.0 & 8.3 \\ \hline
\end{tabular}
\end{center}
\end{table}

\begin{table}[tb]
\caption{\label{tab:bv-test} Comparison of the subsets determined by
the BV analysis according to criteria {\it (i-iii)} with the sites and
diffusion paths identified from the MD simulations. Upper part:
Sensitivities $\psi(\mathcal{P}^{\rm BV}_\star|\mathcal{P}_\star)$;
lower part: specificities $\psi(\mathcal{P}_\star|\mathcal{P}^{\rm
BV}_\star)$.}
\begin{center}
\begin{tabular}
{|@{\hspace{1em}}c@{\hspace{1em}}||@{\hspace{1em}}c@{\hspace{1em}}|@{\hspace{1em}}c@{\hspace{1em}}|@{\hspace{1em}}c@{\hspace{1em}}|}\hline
    & $\mathcal{P}^{\rm BV}_{\rm ex}$& 
$\mathcal{P}^{\rm BV}_{\rm cn}$ &
$\mathcal{P}^{\rm BV}_{\rm ex+cn}$ \\ \hline\hline 
$\mathcal{P}_{\rm sites}$ &
0.83 & 0.95 & 0.79 \\ \hline
$\mathcal{P}_{\rm perc}$ &
0.62 & 0.91 & 0.57 \\ \hline\hline
$\mathcal{P}_{\rm sites}$ &
0.12 & 0.02 &  0.13\\ \hline
$\mathcal{P}_{\rm perc}$ &
0.43 & 0.08 &  0.48\\ \hline
\end{tabular}
\end{center}
\end{table}

\begin{figure}
\includegraphics[width=0.48\textwidth,clip=,]{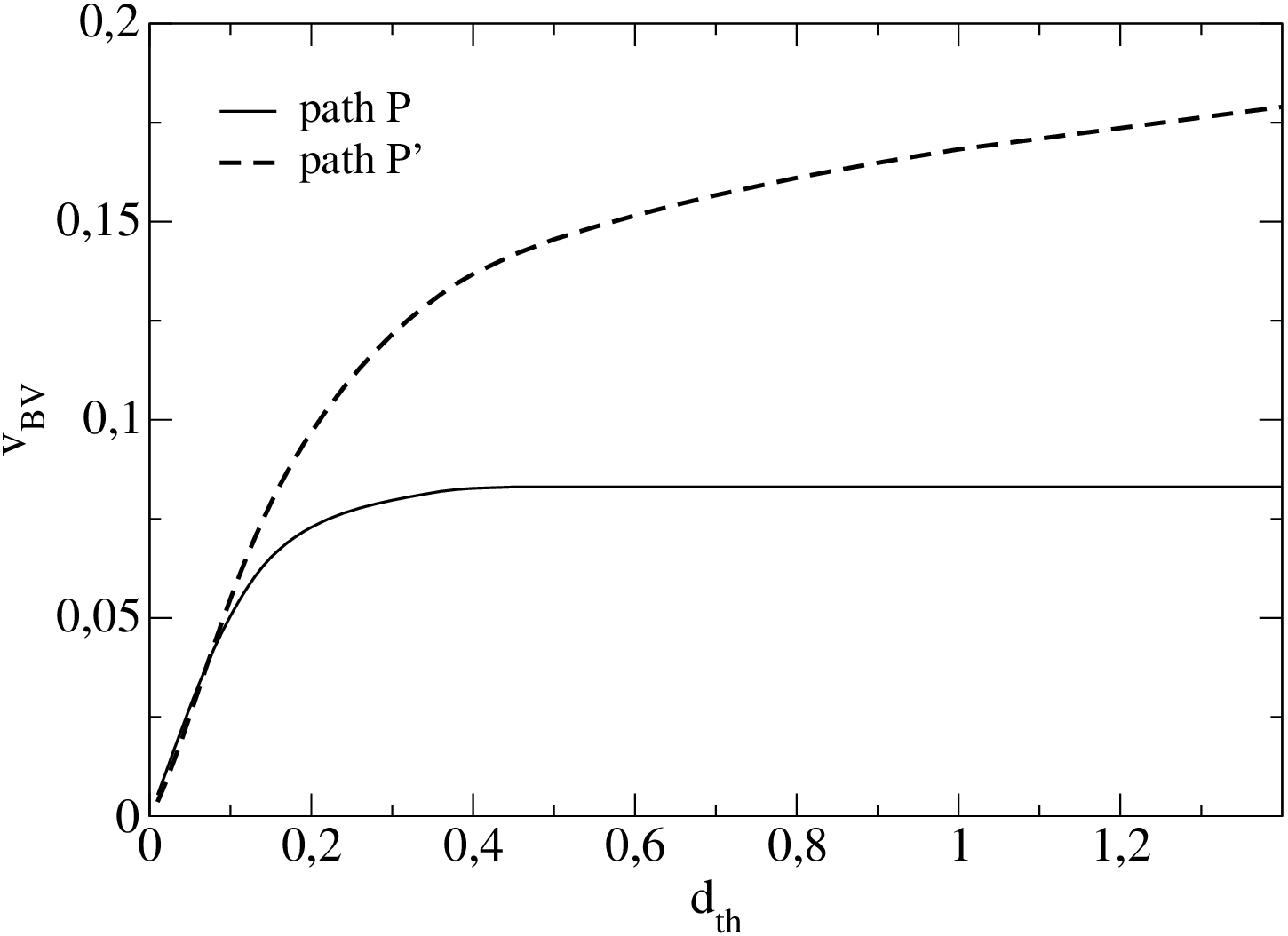}
\caption{\label{fig:bvvolume} Volume fraction $v_{\rm BV}$ of the BV
  path as a functions of the mismatch threshold $d_{\rm th}$.}
\end{figure}

The geometric conditions {\it (i,ii)} already limit the fraction of
available space for the mobile ions to 10\% for the time-averaged
network structure, see table~\ref{tab:volume-fractions}. By contrast,
constraint {\it (iii)} alone for $s_{\rm min}=0.04$ is a weak
condition, which would allow a Li ion to occupy 83\% of the space. The
combined conditions {\it (i-iii)} reduce the accessible space to
8.3\%, corresponding to an uncorrelated behavior. If conditions {\it
(ii-iv)} are replaced by condition {\it (v)} in the limit $d_{\rm
th}\to\infty$ the available space is restricted to 21.3\% of the
volume of the system (see also Fig.~\ref{fig:bvvolume}). One should
expect that this accessible space entails the ionic sites and the
diffusion path identified in sec.~\ref{sec:mdresults}.

To quantify the quality of agreement between the subsets
$\mathcal{P}^{\rm BV}_\star$ ($\mathcal{P}^{\rm
BV}_\star=\mathcal{P}^{\rm BV}_{\rm ex}$, $\mathcal{P}^{\rm BV}_{\rm
cn}$, or $\mathcal{P}^{\rm BV}_{\rm ex+cn}$) with the subsets
$\mathcal{P}_\star$ obtained from the MD simulations
($\mathcal{P}_\star=\mathcal{P}_{\rm sites}$ or $\mathcal{P}_{\rm
perc}$), we define two quantities, the sensitivity and specificity.
The sensitivity is the conditional probability $\psi(\mathcal{P}^{\rm
BV}_\star|\mathcal{P}_\star)$ that a cell belonging to one of the
subsets identified in the MD simulations belongs to one of the subsets
of the BV analysis. Conversely, the specificity of the BV analysis is
quantified by calculating the conditional probabilities
$\psi(\mathcal{P}_\star|\mathcal{P}^{\rm
BV}_\star)$.\cite{specificity-comm}

Table~\ref{tab:bv-test} summarizes the results. Let us in particular
consider the case where conditions {\it (i-iii)} are applied
(corresponding to $\mathcal{P}^{\rm BV}_{\rm ex+cn}$). A high
sensitivity of 79\% is reached for the sites. However, from
table~\ref{tab:volume-fractions} we see that the volume fraction of
$\mathcal{P}^{\rm BV}_{\rm ex+cn}$ (8,3\%) is by a factor of about 6
larger than that of $\mathcal{P}_{\rm sites}$ (1,4\%). Accordingly,
the specificity $\psi(\mathcal{P}_{\rm sites}|\mathcal{P}^{\rm
BV}_{\rm ex+cn})=13\%$ is rather low. The sensitivity for the
diffusion path is 57\% ($\mathcal{P}_{\rm perc}$). It is significantly
lower than that for the sites, since the path contains regions with
very low occupation probability (cf.\ Sec.~\ref{sec:mdresults}), which
often violate the geometric constraints {\it (i,ii)}.

\begin{figure}
\includegraphics[width=0.48\textwidth,clip=,]{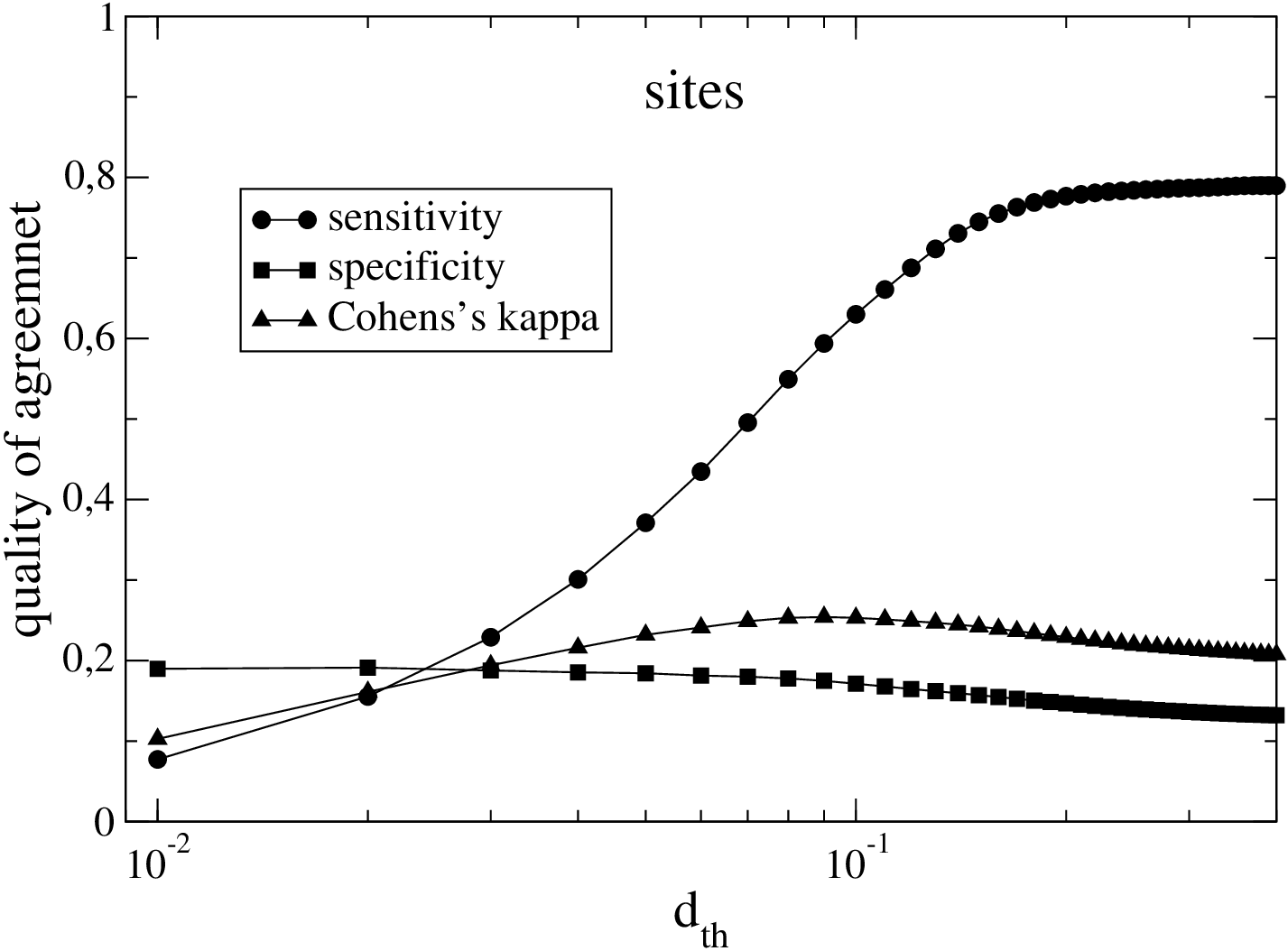}
\includegraphics[width=0.48\textwidth,clip=,]{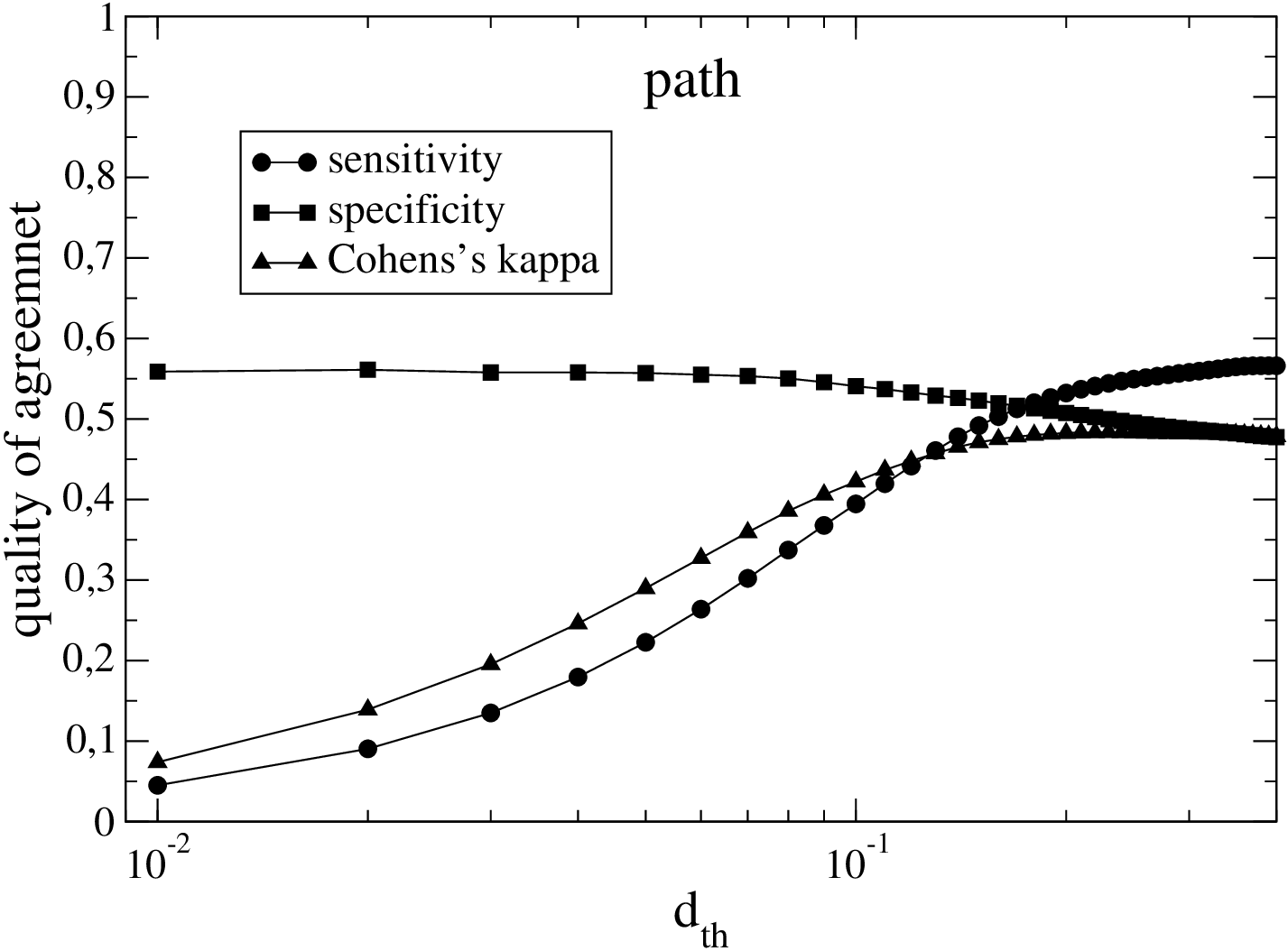}
\caption{\label{fig:sens-spec-bv} Quality of agreement between {\it
(a)} the subsets $\mathcal{P}_{\rm sites}$ and $\mathcal{P}_{\rm
BV}(d_{\rm th})$, and {\it (b)} the subsets $\mathcal{P}_{\rm path}$
and $\mathcal{P}_{\rm BV}(d_{\rm th})$.}
\end{figure}

The question is, whether the specificity can be improved without
significant reduction in the sensitivity by applying condition {\it
(iv)} in addition to {\it (i-iii)}. The volume fraction $v_{\rm BV}$
of the BV path as a function of the threshold mismatch $d_{\rm th}$ is
shown in Fig.~\ref{fig:bvvolume} (solid line). For small $d_{\rm th}$,
$v_{\rm BV}$ increases linearly and for $d_{\rm th}\gtrsim0.3$
approaches the value 8.3\% imposed by the constraints {\it(i-iii)}.
Figures~\ref{fig:sens-spec-bv}a and b show how the sensitivity and
specificity vary with $d_{\rm th}$ with respect to the sites and the
diffusion path, respectively. The sensitivities in
Figs.~\ref{fig:sens-spec-bv}a,b start to saturate for $d_{\rm
th}\gtrsim0.2-0.3$, a value in fair agreement with typical choices
used in BV analyses of RMC models.\cite{Hall/etal:2006} Close to
$d_{\rm th}\simeq0.1$, $v_{\rm BV}\simeq5\%$ is by about 40\% smaller
than the saturation value 8.3\% for $d_{\rm th}\to\infty$ (see
Fig.~\ref{fig:bvvolume}). However, this reduction is not strong enough
to yield a significant improvement of the specificities. Even if one
would take a large loss in the sensitivity by choosing $d_{\rm th}$
very small, the gain in the specificity remains low. In summary we
find that the inclusion of criterion {\it(iv)} does not yield a
substantial improvement and that the essential part of the obtained
agreement is due to the geometric constraints {\it (i,ii)}.

To find an optimal value for $d_{\rm th}$ with respect to both
sensitivity and specificity, we use the maximum of Cohen's kappa
value \cite{Fleiss:1981}. The kappa value for the sets ${\cal A}={\cal
P}^{\rm BV}(d)$ and ${\cal B}={\cal P}_\star$ is defined by
\begin{equation}
\kappa=\frac{(p({\cal A} \cap {\cal B})+ p(\bar{\cal A}\cap \bar{\cal B}))
             - (p({\cal A}) p({\cal B}) + p(\bar{\cal A})p(\bar{\cal B}))}
                  {1-(p({\cal A}) p({\cal B}) + p(\bar{\cal A})p(\bar{\cal B})) },
\label{eq:kappa}
\end{equation}
where $p(.)$ denote the probabilities of the corresponding sets (for
example, $p({\cal A} \cap {\cal B})$ is the probability that a cell
belongs to both the sets ${\cal A}$ and ${\cal B}$). A value of
$\kappa=1$ denotes complete agreement between the two sets, while
$\kappa=0$ corresponds to a random overlap of the two sets (negative
values indicate an anti-correlation). For the sites the maximum of
$\kappa$ occurs at $d_{\rm th}=0.09$ and is only 25\%. This is caused
by the low specificity, which varies only slowly with $d_{\rm th}$. At
$d_{\rm th}=0.09$ we find a specificity of 17\%, while the sensitivity
is 59\%. For the path the maximum of $\kappa$ occurs at $d_{\rm
th}=0.23$ and is 48\%. The sensitivity and specificity attain values
of 54\% and 50\% at this maximum. In summary, the quality of agreement
for the path is promising, while that for the sites is not yet
satisfactory.

\begin{figure}
\includegraphics[width=0.48\textwidth,clip=,]{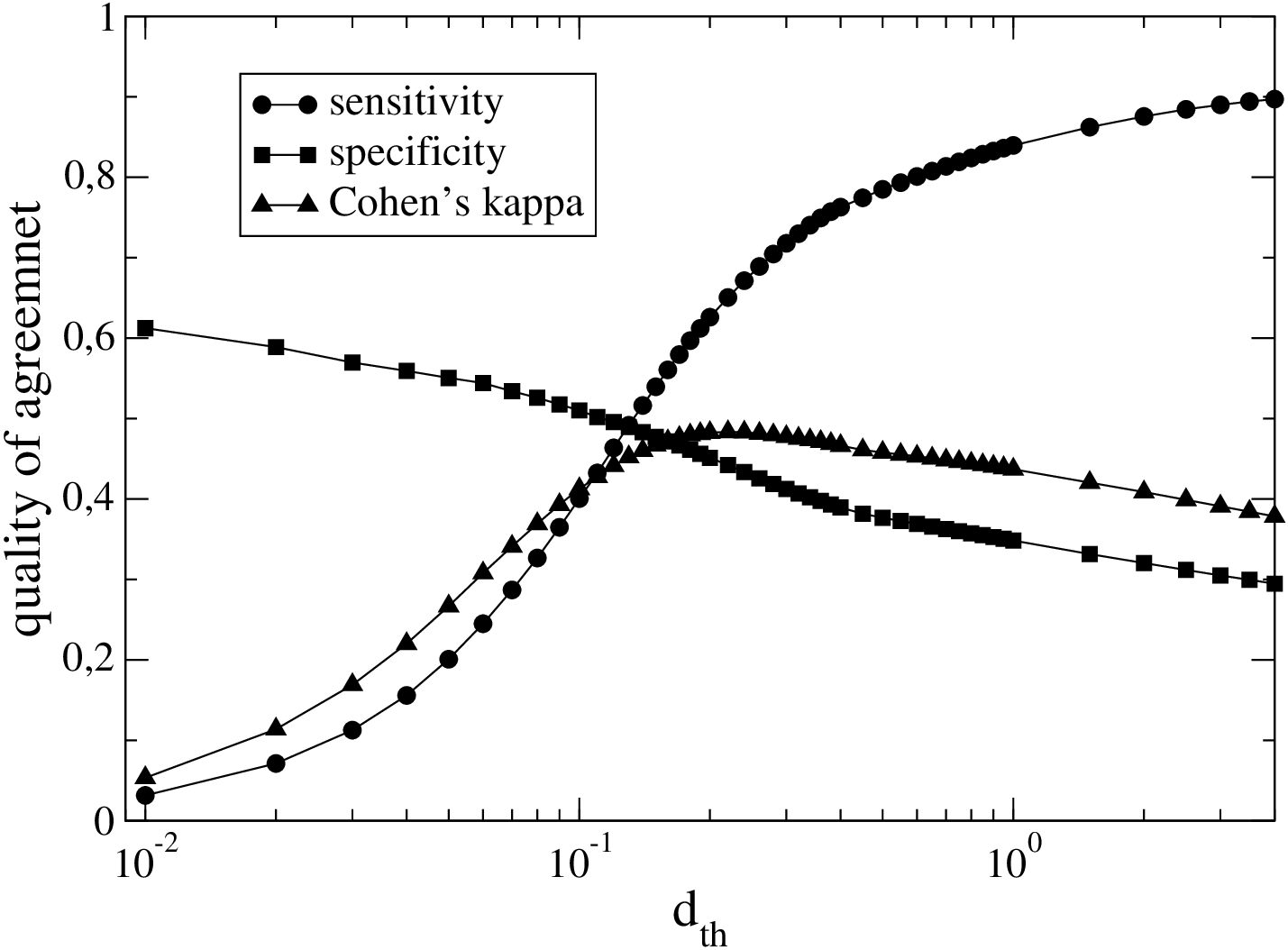}
\caption{\label{fig:sens-spec-kap} Quality of agreement between the
  percolation path ${\cal P}_{\rm perc}$ and $\mathcal{P}_{\rm
  BV}'(d_{\rm th})$ as a function of the BV mismatch threshold $d_{\rm
  th}$.}
\end{figure}

The sensitivity for the diffusion path can however be improved and the
influence of the BV sum mismatch increased by using condition {\it
(v)} instead of conditions {\it (ii-iv)}, i.e. by considering the path
$\mathcal{P}_{\rm BV}'(d_{\rm th})$. For this path in comparison with
$\mathcal{P}_{\rm perc}$ we plot in Fig.~\ref{fig:sens-spec-kap} the
sensitivity and the specificity in dependence of $d_{\rm th}$ (see
eq.~(\ref{eq:cond4b}). For large $d_{\rm th}$, the sensitivity reaches
values up to 90\%, which are significantly higher than the 57\% in
Fig.~\ref{fig:sens-spec-bv}b. The specificity for small $d_{\rm th}$
has values comparable to that in Fig.~\ref{fig:sens-spec-bv}b, and
then decreases to smaller values for larger $d_{\rm th}$. As a
function of $d_{\rm th}$, $\kappa$ reaches a maximum of 48\% at
$d_{\rm th}=0.22$. Further enhancements may be achieved by slight
adjustments of the minimum distance $R_{\rm LiSi}$ or by replacing
criterion {\it (i)} by a penalty function as well, which would mean
that the oversimplifying hard sphere exclusion radius criteria {\it
(i,ii)} may adversely affect the achievable level of agreement.

\begin{figure}
\includegraphics[width=0.48\textwidth,clip=,]{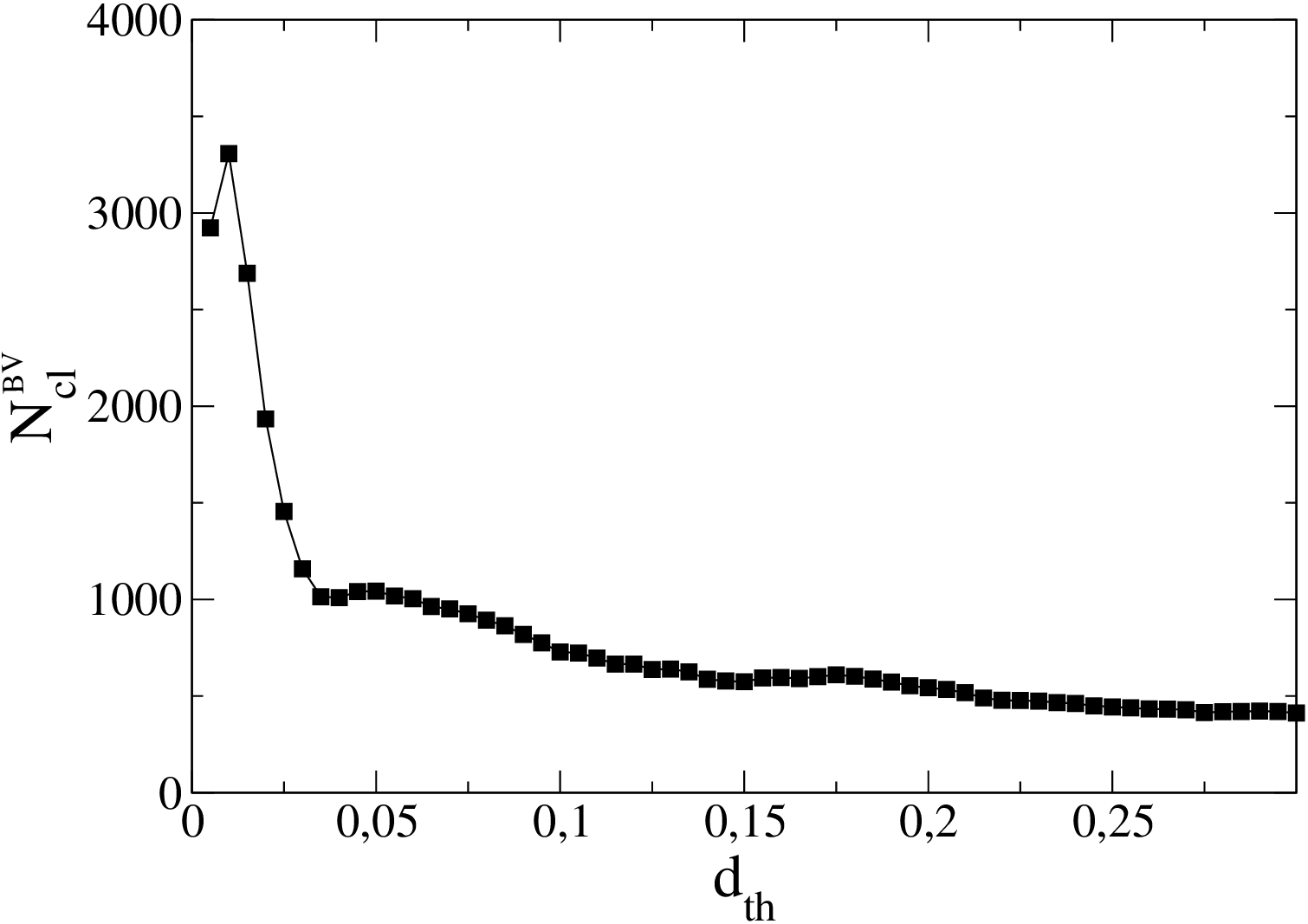}
\caption{\label{fig:ncl-bv} Number of BV clusters as a function of the
  threshold mismatch $d_{\rm th}$.}
\end{figure}

Finally, we test the potential of the BV method to identify sites
(despite the low specificity). To this end we performed a cluster
analysis with respect to $d_{\rm th}$, analogous to the one carried
out in sec.~\ref{sec:mdresults} with respect to $\rho_{\rm cl}$.
Different from the behavior found in Fig.~\ref{fig:nr_of_clusters},
the number $N^{\rm BV}_{\rm cl}$ of BV clusters shown in
Fig.~\ref{fig:ncl-bv} is very large already for small $d_{\rm th}$ and
then decreases rapidly due to coalescence of clusters. Accessible
cells start to percolate at $d_{\rm c}=0.09$. The reason for this
behavior is that the valence mismatch $d(x)$ is a rather rapidly
varying function. Already for small $d_{\rm th}$ there exists a large
number of small disconnected clusters that soon merge together to form
a percolating cluster. While the majority of $v_{\rm BV}$ belongs to
this percolation cluster the fluctuations in $d(x)$ prevent $N^{\rm
BV}_{\rm cl}$ to assume comparatively small values for large $d_{\rm
th}$. Several hundred clusters are found for $d_{\rm th}\simeq0.1$.
Almost all of these are very small: 80\% consist of only a single cell
and only 8 clusters contain more than 100 cells (i.e.\ are comparable
to the size of sites). The percolation cluster includes 95\% of
$v_{\rm BV}(d_{\rm th}=0.1)$. As a consequence the cluster analysis is
not successful for identifying ionic sites with respect to $d_{\rm
th}$.\cite{bv-sites-comm}

\begin{figure}
\includegraphics[width=0.48\textwidth,clip=,]{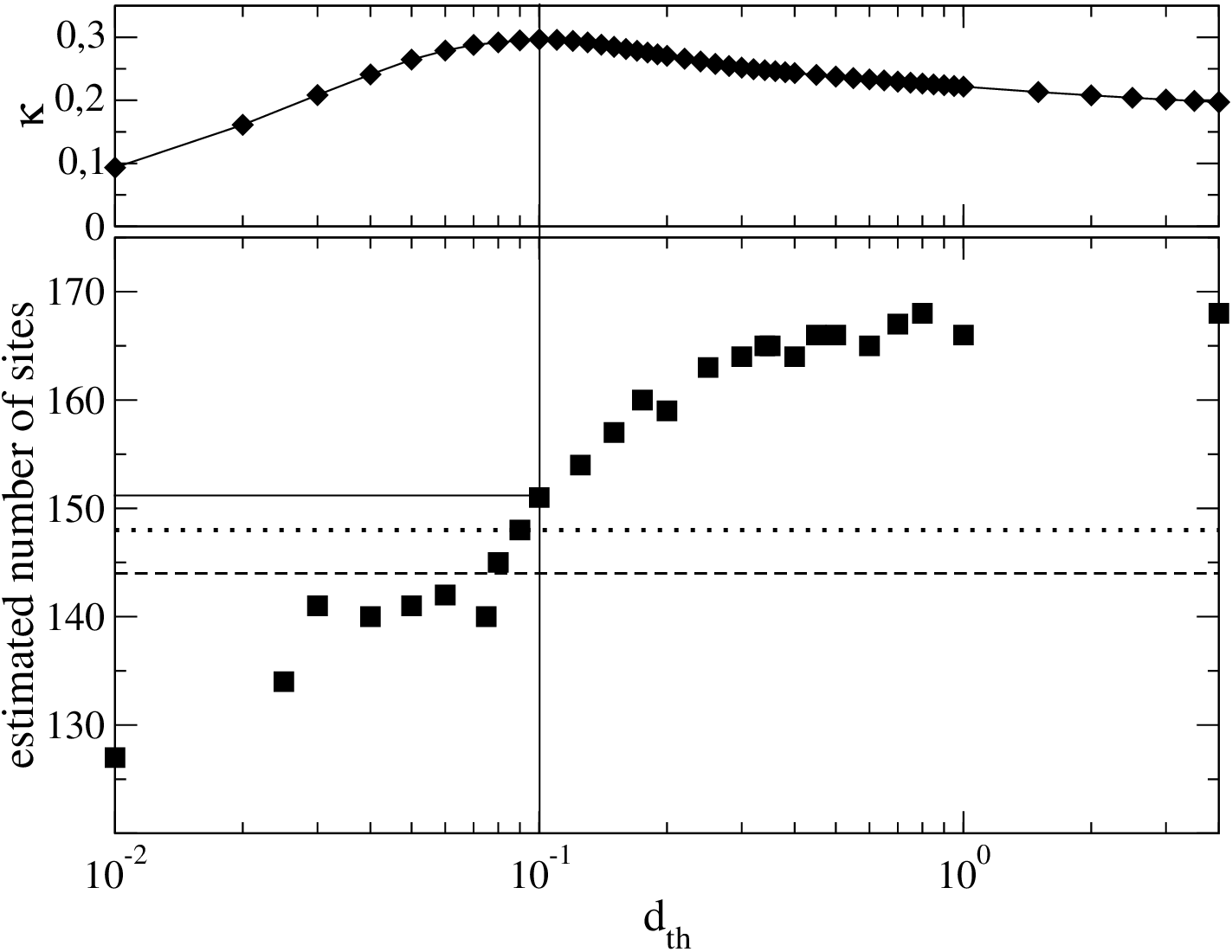}
\caption{\label{fig:site-number} Lower panel: Number of possible site
centers on the BV path $\mathcal{P}_{\rm BV}'(d_{\rm th})$ as
determined by the ``row-by-row algorithm'' described in the text;
Upper panel: Cohen's kappa value for the path $\mathcal{P}_{\rm
BV}'(d_{\rm th})$ in comparison with $\mathcal{P}_{\rm perc}$ (redrawn
from Fig.~\ref{fig:sens-spec-kap}). The number of estimated sites
(151) at the optimal $\kappa$ is indicated by the solid lines. The
dashed line marks the number of Li ions (144) and the dotted line the
number of sites (148) found in Sec.~\ref{sec:mdresults}.}
\end{figure}

It is interesting, however, that the BV analysis can provide a good
estimate of the {\em number} of sites (not their location in space).
It turns out to be useful to include the coordination number
constraint {\it (iii)} in addition to the conditions {\it (i,v)} in
this case. For varying mismatch threshold $d_{\rm th}$ the cells
belonging to the corresponding path $\mathcal{P}_{\rm BV}'(d_{\rm
th})$ are filled row by row under the condition that two centers have
a distance larger than the minimal Li-Li distance $R_{\rm LiLi}=2.62$.
The number of sites found in this way is shown in
Fig.~\ref{fig:site-number} as a function of the mismatch threshold
$d_{\rm th}$ (lower panel). For large $d_{\rm th}$, it approaches the
number of clusters found by the Hoshen/Kopelman analysis in
Sect.~\ref{sec:mdresults}. On the other hand, we can take the
estimated number of sites at the optimal bond valence mismatch $d_{\rm
th}=0.22$ given by the maximum of Cohen's $\kappa$ value with respect
to the sets $\mathcal{P}_{\rm sites}$ and $\mathcal{P}_{\rm
BV}'(d_{\rm th})$. The result displayed in
Fig.~\ref{fig:sens-spec-kap} yields 151 sites. This value is
surprisingly close to the 148 sites found in Sec.~\ref{sec:mdresults}
but studies of further systems are required to test whether such a
prediction is generally possible.

In view of the success of the BV method to estimate the number of
sites, we undertook further attempts to improve the level of agreement
for the localization of the Li sites. If one distributes Li ions more
randomly on the path $\mathcal{P}_{\rm BV}'(d_{\rm th})$ with the same
minimal distance constraint $R_{\rm
LiLi}=2.62$~\AA,\cite{Allen/Tildesley:1987} instead of doing it row by
row as described above, the positioning of centers of sites is no
longer biased to the rim of the BV path. In this way, we obtain 165 BV
clusters for $d_{\rm th}=0.14$. Considering now as BV sites the
spheres with ion radius $R_{\rm LiLi}/2=1.31$~\AA, we found that 75\%
of these sites have at least one cell in common with one of the MD
sites found in Sec.~\ref{sec:mdresults}. Still this is not enough for
a reliable localization of Li sites.

\section{Summary and Perspectives}\label{sec:conclusion}

We have identified sites and diffusion paths for the mobile Li ions in
molecular dynamics simulations of a Li$_2$SiO$_3$ glass. The
identification was based on a cluster analysis of regions with high Li
number density $\rho({\bf x})$. For the clusters to be assigned to
sites, we chose the condition that the total residence time of a Li
ion on it (including intermediate escape to fringe regions) is ten
times larger than the total hopping time to any of its neighboring
clusters. Using this criterion, a very low concentration 2,8\% of
vacant sites was found. An attempt to identify the diffusion path and
sites by bond valence sums calculations was carried out. The
comparison showed that the core of sites and parts of the diffusion
path (up to about 50\%) are captured. Conversely, the BV method was
not suitable to distinguish between regions of high and low $\rho({\bf
x})$, and accordingly not suitable to determine the sites.

When dealing with the MD reference data of this work it proves to be
necessary to adapt the BV parameters derived from experimental
diffraction data to the MD force model. We showed that it is
advantageous to optimize or replace the oversimplifying hard sphere
exclusion radii for application of the BV analysis to the MD model.
Moreover, an improvement of the BV analysis seems to be possible by
adjusting the BV parameters $r_0$ and $\xi$ to the MD force model
instead of only adjusting $V_{\rm id}$ to $\bar V$.

In searching for a powerful method to relate structural properties of
the disordered network structure to transport properties of the mobile
ions a further approach could be based on an effective potential
$U_{\rm eff}({\bf x})\propto-k_{\rm B}T\ln\rho({\bf x})$. Taking this
effective potential one could perform a critical path analysis to
determine the activation energy for the long-range ion mobility (see
e.g.\ ref.~\onlinecite{Maass:1999}). This approach would incorporate
Coulomb interaction effects between the mobile ions in a mean field
type approximation. Preliminary studies by us show that such Coulomb
effects are important for the formation of the sites. With respect to
the BV method we have also performed preliminary studies to evaluate
the degree of correlation between $U_{\rm eff}({\bf x})$ and the BV
mismatch $d({\bf x})$ (or $d'({\bf x})$). It turned out that BV method
cannot clearly distinguish between regions of high and low $\rho({\bf
x})$ and accordingly it is not suitable to locate the sites. Besides
the described shortcomings of a particular choice of parameter values
this is mainly caused by neglecting the long-range Li-Li Coulomb
repulsions.

The precise decomposition of the ion trajectories into residence and
transition parts described in Sec.~\ref{sec:mdresults} moreover may
allow one to bridge the time scale gap between molecular dynamics and
coarse-grained Monte-Carlo simulations. Given two neighboring sites,
many transitions of mobile ions can be followed and the average
transition rate calculated. The elementary rates for the possible
transitions can then be used in a subsequent Monte-Carlo simulation.
However, such procedure is not particularly useful if one is not able
to take into account the temperature dependence of the elementary
rates. One possibility is to use MD data at various high temperatures,
and to try to calculate from them activation energies for the
elementary rates (if these follow an Arrhenius type behavior). With
less effort, one could use again the density $\rho({\bf x})$ and the
effective potential derived from it to determine the energetics of a
coarse-grained hopping model. Further investigations in this direction
have to be undertaken in order to establish a working multi-scale
modeling of these complex amorphous systems.

\section*{Acknowledgments}
Financial support to S.~A. by the NUS ARF (R-284-000-029-112/133) and
to the authors from TU Ilmenau by the HI-CONDELEC EU STREP project
(NMP3-CT-2005-516975) is gratefully acknowledged.

\newpage

\begin{thebibliography}{22}
\bibitem{vacancy-comm} We note that the vacancies are generated as
  defect structures in the glassy network, which becomes frozen below
  the glass transition. Accordingly, their number concentration does
  not change with temperature (it may be slightly affected by aging
  effects). The situation is thus different from that in crystalline
  systems, where the number concentration of vacancies is in general
  thermally activated.
\bibitem{dispersive-comm} The $r^{-6}$ dependence can be motivated by
  the van-der-Waals interaction. However, since this term is used both
  for distances corresponding to covalent bonding as well as large
  distances, it in fact is not a van-der Waals interaction and often
  referred to as ``dispersive interaction''. Nevertheless, the
  amplitude factors $c_i$, $c_j$ are usually assumed to scale with the
  polarizabilities. This gives reason to include this term here only
  for the large oxygen ions.
\bibitem{time-comm} The total hopping and residence time, of course,
  have to be normalized with respect to the simulation time to give an
  intrinsic cluster property. Alternatively, we could have defined
  averaged (with respect to the number of events) hopping and
  residence times. In order to calculate the occupation probabilities
  in Fig.~\ref{fig:res-hop-time}a we preferred to determine the total
  times.
\bibitem{specificity-comm} This definition of specificity deviates
  from the usual one, as it is, for example, often applied in medical
  studies. In such studies the specificity is given by the conditional
  probability $\psi(\bar{\mathcal{P}}^{\rm
  BV}_\star|\bar{\mathcal{P}}_\star)$ with $\bar{\mathcal{P}}^{\rm
  BV}_\star$ and $\bar{\mathcal{P}}_\star$ being the complements of
  $\mathcal{P}^{\rm BV}_\star$ and $\mathcal{P}_\star$, respectively.
  Both definitions are related by
  $\psi(\mathcal{P}_\star|\mathcal{P}^{\rm
  BV}_\star)=\psi(\mathcal{P}_\star)\psi(\mathcal{P}^{\rm
  BV}_\star|\mathcal{P}_\star)/[\psi(\mathcal{P}_\star)\psi(\mathcal{P}^{\rm
  BV}_\star|\mathcal{P}_\star)+(1-\psi(\mathcal{P}_\star))(1-\psi(\bar{\mathcal{P}}^{\rm
  BV}_\star|\bar{\mathcal{P}}_\star))]$, where
  $\psi((\mathcal{P}_\star)$ is the probability that a cell belongs to
  $\mathcal{P}_\star$ (i.e.\ equal to the volume fraction given in
  table~\ref{tab:volume-fractions}).
\end{thebibliography}
\end{document}